\documentclass[twocolumn]{aastex63}

\usepackage{amstext}
\usepackage{amsmath}
\usepackage{amssymb}
\usepackage{graphicx}

\newcommand{\beq}{\begin{equation}}
\newcommand{\eeq}{\end{equation}}
\newcommand{\beqar}{\begin{align}}
\newcommand{\eeqar}{\end{align}}

\newcommand{\Mbh}{M_{\rm bh}}
\newcommand{\Ms}{M_\ast}
\newcommand{\Rs}{R_\ast}
\newcommand{\rM}{r_{\rm m}}
\newcommand{\rh}{r_{\rm h}}
\newcommand{\rp}{r_{\rm p}}
\newcommand{\rt}{r_{\rm t}}
\newcommand{\E}{{\cal E}}

\shorttitle{}

\shortauthors{}

\begin{document}

\title{The Effect of Star--Disk Interactions on Highly Eccentric Stellar Orbits in Active Galactic Nuclei: A Disk Loss Cone and Implications for Stellar Tidal Disruption Events}

\author[0000-0002-1417-8024]{Morgan MacLeod}
\affiliation{Harvard-Smithsonian Center for Astrophysics, 60 Garden Street, Cambridge, MA, 02138, USA}
\email{morgan.macleod@cfa.harvard.edu}

\author{Douglas N. C. Lin}
\affiliation{Department of Astronomy \& Astrophysics, University of California, Santa Cruz, CA, 95064, USA}

\begin{abstract}
Galactic center black holes appear to be nearly universally surrounded by dense stellar clusters. When these black holes go through an active accretion phase, the multiple components of the accretion disk, stellar cluster, and black hole system all coexist. We analyze the effect of drag forces on highly eccentric stellar orbits incurred as stars puncture through the disk plane. Disk crossings dissipate orbital energy, drawing eccentric stars into more circular orbits. For high surface density disks, such as those found around black holes accreting near the Eddington mass accretion limit, the magnitude of this energy dissipation can be larger than the mean scatterings that stars receive by two body relaxation. One implication of this is the presence of a disk ``loss cone" for highly eccentric stellar orbits where the dissipation from disk interaction outweighs scatter via two body relaxation. The disk loss cone is larger than the tidal disruption loss cone for near-Eddington black hole accretion rates. Stellar orbits within the disk loss cone are lost from the overall cluster as stellar orbits are circularized and stars are potentially ablated by their high-velocity impacts with the disk. We find, however, that the presence of the disk loss cone has a minimal effect on the overall rate of stellar tidal disruptions. Stars are still efficiently fed to the black hole from more-distant stellar orbits that receive large-enough per orbit scatter to jump over the disk loss cone and end up tidally disrupted. 
\end{abstract}


\section{Introduction}
Dense stellar clusters surround and coexist with supermassive black holes in the centers of galactic nuclei \citep[e.g.][]{2013ARA&A..51..511K}. These clusters are the product of dynamical relaxation of stellar orbits in the combined potential of the black hole and surrounding stars \citep{2013degn.book.....M}. When gas is funneled into these nuclear regions, it can lead to the assembly of accretion disks, which transport material toward the supermassive black hole.  The associated accretion luminosity makes these black holes visible as active galactic nuclei \citep[AGN, e.g.][]{1993ARA&A..31..473A,1996ARA&A..34..703L} 

Within this complex system of black hole, stars, and gaseous accretion flow, close-in stellar orbits interact continuously with the accretion disk surrounding the black hole \citep[e.g.][]{1983ApJ...273...99O,1991MNRAS.250..505S,1993ApJ...409..592A}. While the stellar distribution is thought, in many cases, to be largely isotropic, the accretion structure may form a thin, equatorial disk. Stellar orbits that are inclined relative to the disk pass through the disk twice per orbit, leading to drag forces as the stars  interact with the disk gas. 

Besides the simple fact of ongoing star--disk interactions, previous work has focused on several different aspects of this complex problem. The orbits of stars themselves are modified by disk crossings \citep{ 1983ApJ...273...99O,1991MNRAS.250..505S,1993MNRAS.265..365V,1995MNRAS.275..628R, 1998MNRAS.293L...1V,1998MNRAS.298...53V,2001A&A...376..686K} and torques from the disk's gravitational potential \citep{1993ApJ...409..592A, 1998MNRAS.298...53V, 2005A&A...433..405S}. The collective effect of these encounters has the potential to modify the stellar distribution around an accreting black hole \citep{1991MNRAS.250..505S,1995MNRAS.275..628R,2004MNRAS.354.1177S,2005ApJ...619...30M,2016MNRAS.460..240K,2018MNRAS.476.4224P}. 

Additionally, significant effort has gone into understanding the effects of disk-crossing impacts on stars themselves and their structures. For example, depending on their stellar type, stars and compact objects might accrete \citep{1993ApJ...409..592A,1994ApJ...423L..19L,1995MNRAS.276..597R,1998ApJ...507..131I, 2011MNRAS.417L.103M,2012MNRAS.425..460M} or might be stripped by the ram pressure of disk-crossing impacts \citep{ 1994ApJ...434...46Z,1996ApJ...470..237A,2016ApJ...823..155K}. Relatedly, star--disk crossings might impact disk structure \citep{1983ApJ...273...99O,1994ApJ...434...46Z,1995MNRAS.276..597R,1998ApJ...507..131I,2006IJMPD..15.1001D,2006Ap&SS.305..187D,2007ApJ...658..114P,2012ApJ...748...63B}, or lead to the feeding of stars and stellar material to the black hole \citep{1993ApJ...409..592A,1993MNRAS.264..388K,1994A&A...292..404H,1996ApJ...470..237A,2005ApJ...619...30M,2006NewAR..50..786M,2007A&A...470...11K,2012EPJWC..3901003K,2013MNRAS.434.2948D,2016MNRAS.460..240K,2017ARA&A..55...17A}. As gravitational-wave observations of merging compact objects have become possible, renewed focus on this field has come through discussion of AGN disks as possible sites of enhanced merger rates of binary compact objects \citep[e.g.][]{2011PhRvD..84b4032K,2016ApJ...819L..17B,2017MNRAS.464..946S,2017ApJ...835..165B,2018ApJ...866...66M,2019ApJ...878...85S, 2019arXiv190704871F,2019arXiv190704356M,2019arXiv190703746M,2019ApJ...876..122Y}.

Here we examine the question of how star-disk interactions affect  the highest eccentricity stellar orbits that pass close to the supermassive black hole at periapse \citep[e.g.][]{2005A&A...433..405S}. This question is of relevance to the possible coexistance of tidal disruption events of stars with AGN accretion flows. One such potential event has been observed by \citet{2017NatAs...1E..61T}, who additionally claim that this event also represents a highly elevated occurrence rate of tidal disruptions within AGN. Theoretical models of the hydrodynamic interaction between a tidal disruption debris stream and a preexisting accretion disk have been recently presented by \citet{2019ApJ...881..113C}. 

To examine this question, we adopt simplified models of accretion disk structure, stellar orbital properties, and the consequences of star-disk interaction, which we describe in Section \ref{sec:model}. In Section \ref{sec:results}, we describe our results and present an order-of-magnitude model which captures the main features of orbital modifications due to disk crossings. In Section \ref{sec:stardisk}, we discuss potential consequences of our findings for the coexistance of stars and accretion disk structures in galactic nuclei, in particular the depletion of eccentric orbits and potential implications for tidal disruption events. In Section \ref{sec:conclusion}, we conclude.

\section{Star-Disk Interaction Model}\label{sec:model}

In this section, we outline our model for a nuclear stellar cluster and accretion disk surrounding a central supermassive black hole in an AGN. The following methods are encapsulated in a publicly released python package {\tt NSC\_dynamics}, which is released along with examples and tests.\footnote{url: \url{github.com/morganemacleod/NSC_dynamics} }

\subsection{Nuclear Star Cluster}\label{sec:nsc}
We model the stellar cluster surrounding the supermassive black hole of mass $\Mbh$ as follows. We adopt the simplification that stars share a single stellar mass, $\Ms$, and stellar radius, $\Rs$, and a spherical, power law stellar distribution.

The stellar number density is arranged according to a power-law distribution in radius, 
\beq
\nu_\ast = \nu_{\rm m}  \left(\frac{r}{\rM}\right)^{-\gamma},
\eeq 
with $\gamma=1.5$, and where $\rM$ is the radius that encloses twice the black hole mass in stars  \citep[equation 2.11]{2013degn.book.....M}.  The normalization \citep[][equation 3.48]{2013degn.book.....M},
\beq
\nu_{\rm m}=\frac{3-\gamma}{2\pi} \frac{\Mbh}{\Ms} \rM^{-3},
\eeq
assures that the integrated stellar mass is $2\Mbh$ within $\rM$.
To set the length scale of $\rM$, we turn to the black hole mass--velocity dispersion relation, which implies that the typical velocity dispersion is 
\beq
\sigma_{\rm h} = 2.3 {\rm \ km \ s}^{-1} (\Mbh/M_\odot)^{1/4.38} ,
\eeq
with the numerical values derived from the fit of \citet{2013ARA&A..51..511K}.  This means that the black hole is the dominant influence on stellar motions within an influence radius,
\beq
\rh= \frac{G\Mbh}{\sigma_{\rm h}^2},
\eeq
 where we follow the notation of \citet[equation 2.12]{2013degn.book.....M}. 
We will assume that that $\rM =  \rh$, thus setting the normalization of the stellar density profile.  

We adopt the Keplerian limit in which the black hole dominates the gravity in which stars orbit. The orbital period is, therefore, $P = 2\pi \left(r^3 / G\Mbh \right)^{1/2}$ for orbits of semi-major axis $a=r$. 
If the stellar distribution function is isotropic in angular momentum space, the distribution function depends on specific energy (which is defined positive), 
\beq
\E=\frac{G\Mbh}{2a},
\eeq
only, and simplifies to a power law, 
\beq\label{fE}
f(\E) = f_0 {\E}^{\gamma - {3 \over 2}},
\eeq
where 
\beq
f_0 = (2\pi)^{-{3\over2}} \nu_{\rm m}  \Phi_0^{-\gamma} \frac{\Gamma(\gamma+1)}{\Gamma(\gamma-{1\over2})}, 
\eeq
in which $\Phi_0 = G \Mbh / r_m$ \citep{2013ApJ...774...87V}.

The three-dimensional stellar velocity dispersion for such a structure is approximately,
\beq
\sigma^2 \approx \frac{G \Mbh}{(1+\gamma)r}.
\eeq
The two-body stellar relaxation time is 
\beq\label{eq:trel}
t_{\rm rel} \approx \frac{0.34 \sigma^3}{G^2 \Ms^2 \nu_\ast\ln\Lambda},
\eeq
where $\ln\Lambda\approx\ln\left(\Mbh/\Ms\right)$ is the Coulomb logarithm \citep{2013degn.book.....M}.  Over this timescale, the energy and angular momentum of stellar orbits in the cluster are randomized by the cumulative effect of two-body scatterings. Over an orbit, the magnitude of the typical root-mean-square (RMS) change in energy is 
\beq\label{eq:DeltaErms}
\Delta \E_{\rm rel} \approx \E \left( \frac{P}{t_{\rm rel}} \right)^{1/2}, 
\eeq
where $P \approx 2\pi \left(r^3 / G \Mbh\right)^{1/2}$ is the orbital period of a stellar orbit with semi-major axis $r$. Similarly, the magnitude of the RMS change in angular momentum is 
\beq\label{eq:DeltaJrms}
\Delta J_{\rm rel} \approx J_{\rm c} \left( \frac{P}{t_{\rm rel}} \right)^{1/2}, 
\eeq
where $J_{\rm c}=\left(G \Mbh r \right)^{1/2}$ is the circular angular momentum for semi-major axis equal to $r$.

\subsection{AGN disk}

\begin{figure}[htbp]
\begin{center}
\includegraphics[width=0.48\textwidth]{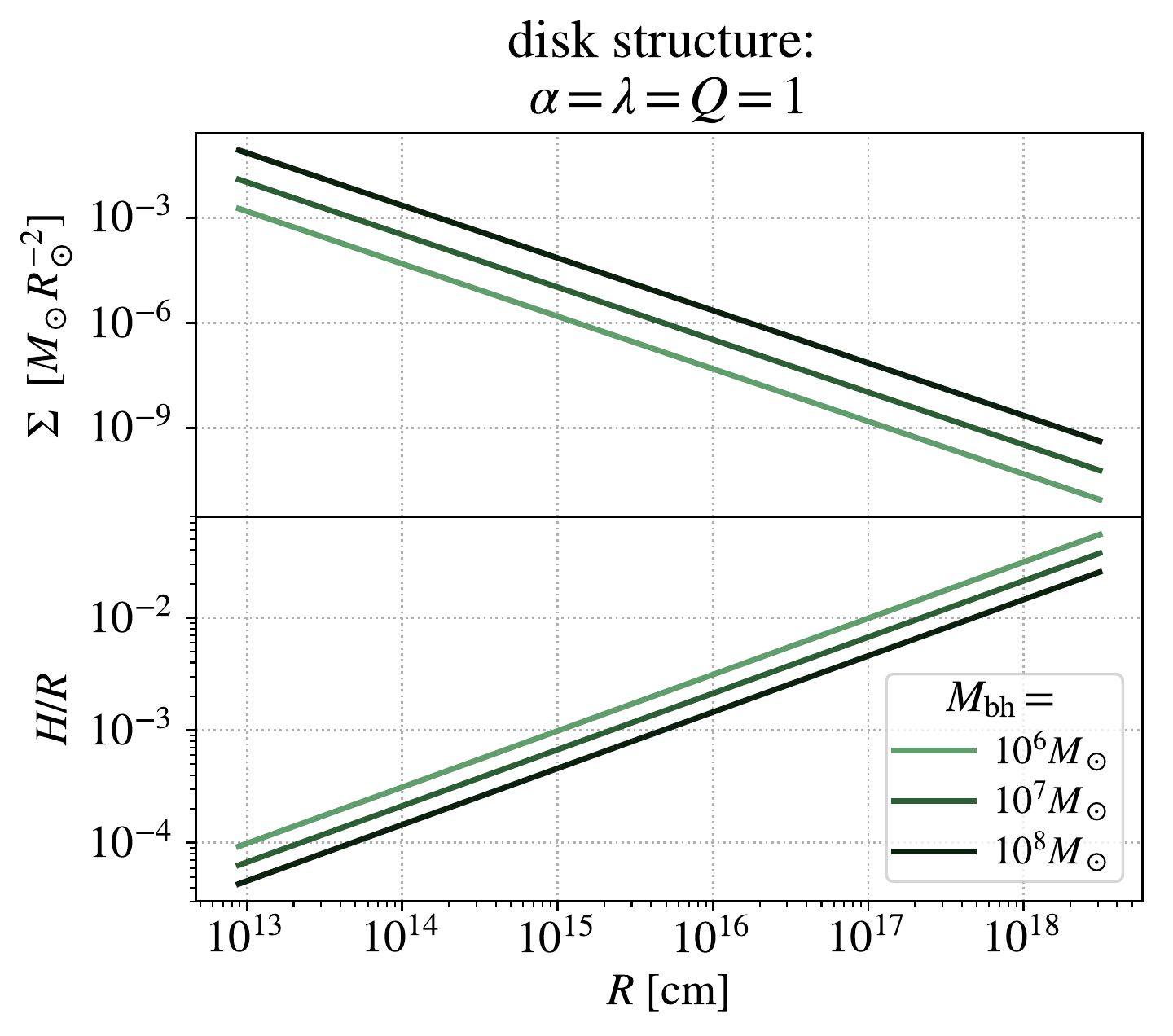}
\caption{Example disk structures in the model adopted. Disk surface densities (top panel) scale with Toomre $Q$ and the critical surface density, equation \eqref{sigma_crit}. The dimensionless scale height, $h=H/R$, is an increasing function of $R$. The panels above are relevant to a disk that is at the threshold of gravitational stability everywhere, $Q=1$, that is accreting onto the black hole at the Eddington limit, $\lambda=1$, where we presume that $\alpha\sim1$ is a result of gravitational instability.    }
\label{fig:disk}
\end{center}
\end{figure}

An accretion disk coexists with the nuclear stellar cluster in the AGN nucleus. Here we describe our disk model. 

We adopt a disk model in which the disk has constant mass flux, $\dot M$, constant Toomre $Q$ parameter, describing its susceptibility to gravitational instability, and constant $\alpha$, parameterizing the efficiency of instability-driven ``viscosity" in the disk \citep[for more description of the role of $\alpha$, see, for example, the review of][]{1995ARA&A..33..505P}. 

We parameterize the disk accretion rate in terms of the Eddington mass accretion rate, $\dot M_{\rm Edd} = L_{\rm Edd}/\eta c^2$, or 
\beq\label{eq:eddington}
\dot M = \lambda \dot M_{\rm Edd} =  0.22~\lambda \left( \frac{\Mbh}{10^7 M_\odot}\right) M_\odot~{\rm yr}^{-1},
\eeq
where we have adopted $\eta=0.1$ and used $L_{\rm Edd} = 4\pi G \Mbh m_{\rm p} c / \sigma_{\rm es}$, in which $m_{\rm p}$ is the proton mass and $ \sigma_{\rm es}$ is the Thompson cross section to electron scattering. 

The disk surface density is given by the mass flux and the radial velocity of material through the disk,
\beq
\Sigma = \frac{\dot M}{2\pi r v_r},
\eeq
where the radial velocity, $v_r = \alpha h^2 (G \Mbh / r)^{1/2}$, and where $h=H/R$ is the dimensionless disk scale height \citep{1995ARA&A..33..505P}. Under these assumptions, the average volume density within the disk is approximately,
\beq
\rho \approx {\Sigma \over 2H} .
\eeq
The dimensionless scale height, $h$, is given by the relation 
\beq
h^3 \approx {Q\over 2 \alpha}{\dot M \over \Mbh \Omega } ,
\eeq
where $\Omega = \left(G \Mbh/r^3 \right)^{1/2}$ is the disk angular frequency. Under these conditions, the surface density can also be written in terms of the critical surface density for gravitational instability, 
\beq\label{sigma_crit}
\Sigma = {\Sigma_{\rm c} \over Q } = {h \over Q} {\Mbh \over \pi r^2} = \frac{1}{(2\alpha)^{1/3}Q^{2/3}} {\Mbh^{2/3} \dot M^{1/3} \over \pi r^2 \Omega^{1/3}}. 
\eeq
Thus the scaling of $\Sigma$ with disk parameters is $\Sigma \propto \alpha^{-1/3} \lambda^{1/3} Q^{-2/3}$.  Figure \ref{fig:disk} shows example disk profiles for the model in which $\alpha=\lambda = Q=1$, which we consider in Section \ref{sec:results}. 

We adopt an orientation such that the disk is in the $X-Y$ plane, with angular momentum in the $+Z$-direction. The disk velocity, $\vec V_{\rm disk}$ at position $\vec X$ is set by the circular, Keplerian velocity at cylindrical radius $R= \sqrt{X^2 + Y^2}$. 

\subsection{Stellar Orbits}
We define stellar orbits on the basis of the orbital elements: semi-major axis, $a$, eccentricity, $e$, longitude of the ascending node, $\Omega$, argument of periapsis, $\omega$, inclination, $I$, and true anomaly, $f$. Together this set of elements allows complete specification of the orbital position and velocity \citep{1999ssd..book.....M}.  

The ranges of the orbital elements are $0<a<\infty$ and $e\leq 0 < 1$, implying bound orbits, $-\pi < \Omega \leq \pi$,  $-\pi < \omega \leq \pi$, and $0 \leq I < \pi$. The magnitude of the specific orbital energy is $\E$, and the orbital vector angular momentum is 
\beq\label{eq:Jvec}
\vec J = \vec{X}_\ast \times \vec{V}_\ast,
\eeq
where $\vec{X}_\ast$ and $\vec{V}_\ast$ are the stellar position and velocity in the reference frame. 
Inclinations $0 \leq I < \pi/2$ are ``prograde", implying that orbit and disk angular momentum vectors both have positive $Z$-components, while orbits $\pi/2 <  I < \pi$ are ``retrograde". 

\begin{figure}[tbp]
\begin{center}
\includegraphics[width=0.48\textwidth]{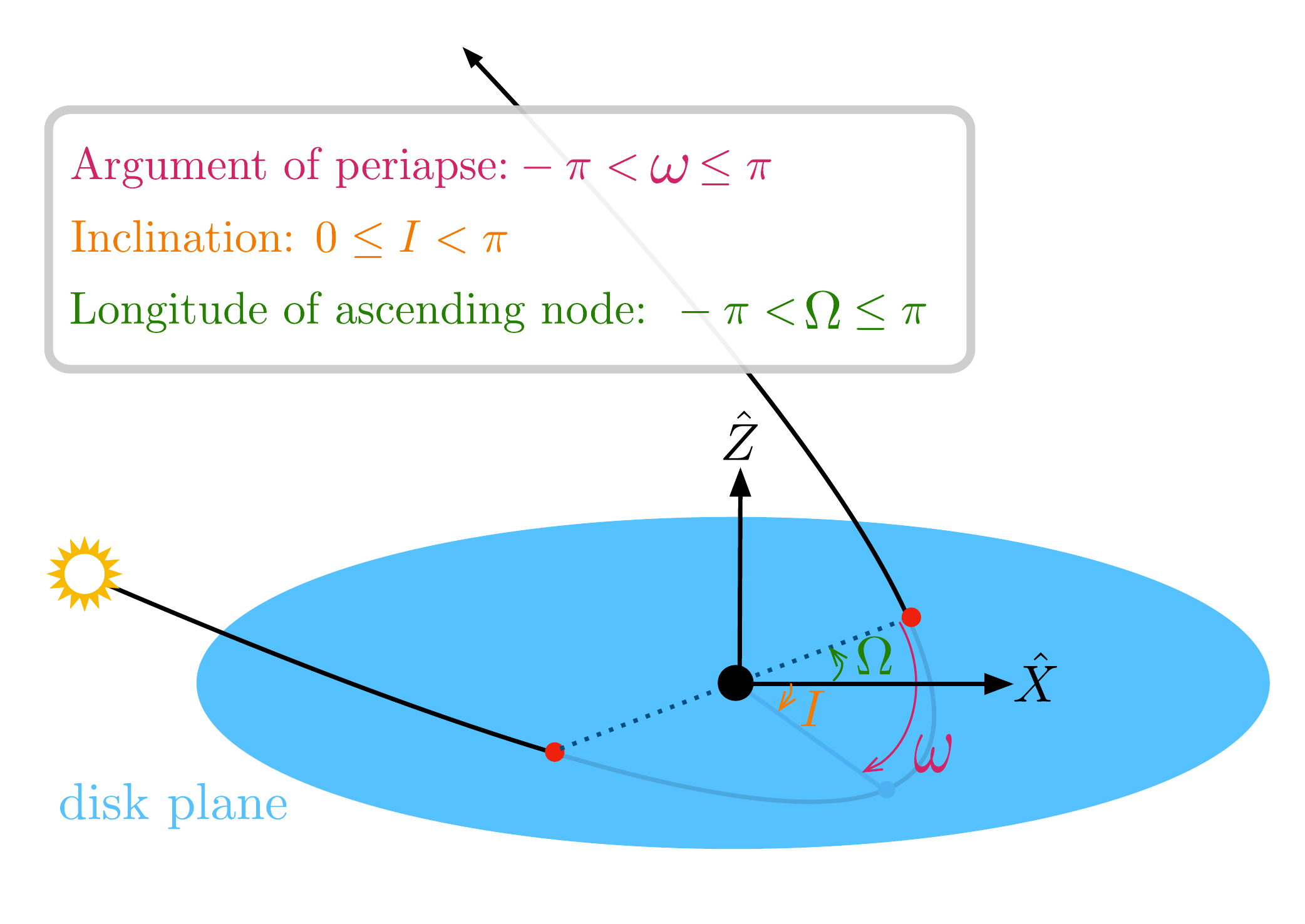}
\caption{Schematic highlighting the interaction of an eccentric stellar orbit with close-in regions of the AGN accretion flow. Here the orbit crosses the disk plane (defined by angular momentum in the $+\hat Z$-direction) at two nodes marked with red dots. Orbits are described by semi-major axis, $a$, eccentricity, $e$, and orientation through three angles: $I$, $\omega$, and $\Omega$, labeled above. In this example, $\Omega>0$ and $\omega<0$.   }
\label{fig:diagram}
\end{center}
\end{figure}

\subsection{Star-Disk Interactions}

 We will work in the approximation that the disk is thin ($h\ll1$), in which case disk crossings occur over a small arc length of the orbit near the nodes. We will, therefore, consider positions and relative velocities to be those evaluated at the disk midplane or node crossings, but we note that this approximation would not be appropriate for a thick disk.
Because we are interested in positions and velocities of node crossings, in which the stellar orbit transects the disk plane, we begin with the true anomaly of node crossings, $f = -\omega$ and $f=-\omega+\pi$. Given these we compute positions and velocities in the plane of the individual stellar orbit, then rotate these to the reference frame. 

Node crossings correspond to $Z=0$ in the reference frame. Using the orbital elements to compute the position $\vec{X}_\ast$ and velocity, $\vec{V}_\ast$, of the star at node crossings, we then apply the drag force resulting from relative motion between the star and the disk.  
The relative velocity is
\beq
\vec{V}_{\rm rel}= \vec V_\ast - \vec V_{\rm disk}.
\eeq
Given the magnitude of a drag force, $F_{\rm drag}$, the change in orbital velocity, in the impulse approximation, is 
\beq\label{eq:DeltaV}
\Delta \vec{V} = - {F_{\rm drag} \over \Ms}  \Delta t_{\rm cross} { \vec{V}_{\rm rel} \over |\vec{V}_{\rm rel}| },
\eeq
in which the crossing time through the disk is
\beq\label{eq:tcross}
\Delta t_{\rm cross} = {2H\over V_{\ast,Z}}.
\eeq

Finally, the drag force may be set by either the gravitational cross section or geometric cross section of the star, depending on the relative velocity of the star through the gas. To order of magnitude, these forces are,
\beq\label{eq:Fgrav}
F_{\rm drag,grav} \approx { 4\pi \left( G \Ms \right)^2  \rho \over   |\vec{V}_{\rm rel}|^2},
\eeq
and 
\beq\label{eq:Fgeo}
F_{\rm drag,geo}  \approx \pi \Rs^2  \rho  |\vec{V}_{\rm rel}|^2. 
\eeq
The resultant drag force is the maximum of these two,
\beq\label{eq:Fdrag}
F_{\rm drag} = {\rm max} \left( F_{\rm drag,grav}, F_{\rm drag,geo} \right),
\eeq
where, in general, the geometric cross section dominates when $|\vec{V}_{\rm rel}| \gg \left(G\Ms/\Rs\right)^{1/2}$ and the gravitational cross section is important when $|\vec{V}_{\rm rel}| \ll \left(G\Ms/\Rs\right)^{1/2}$.

\section{Results}\label{sec:results}

To understand the effects of disk crossings on eccentric orbits, we base most of our analysis on a representative system consisting of a black hole of mass $\Mbh = 10^7 M_\odot$, surrounded by a disk with $\dot M = \dot M_{\rm Edd}$, ($\lambda=1$, equation \ref{eq:eddington}).  We assume that this disk has the critical surface density, such that Toomre's $Q=1$ everywhere, and adopt $\alpha=1$ throughout as well. Such a model is representative of the high-accretion rate extremes of AGN activitiy, in which gravitational instability, and associated non-linear instabilities self-regulate to ensure that $Q$ and $\alpha$ are both of order unity \citep{1995ARA&A..33..505P}.

In Section \ref{sec:modification}, we describe the basic modifications experienced by highly-eccentric orbits by disk interactions. We compare the magnitude of these changes to two-body relaxation in Section \ref{sec:twobody}. In Section \ref{sec:oom}, we provide an analytic framework for interpreting these results and discuss their scaling to other black hole masses, accretion states, and stellar orbit properties. Section \ref{sec:co} extends these results to consider the case of compact objects, where gravitational focus rather than geometric size determines the encounter cross section and drag force.

\subsection{Modification of Eccentric Orbits}\label{sec:modification}

\begin{figure*}[tbp]
\begin{center}
\includegraphics[width=0.9\textwidth]{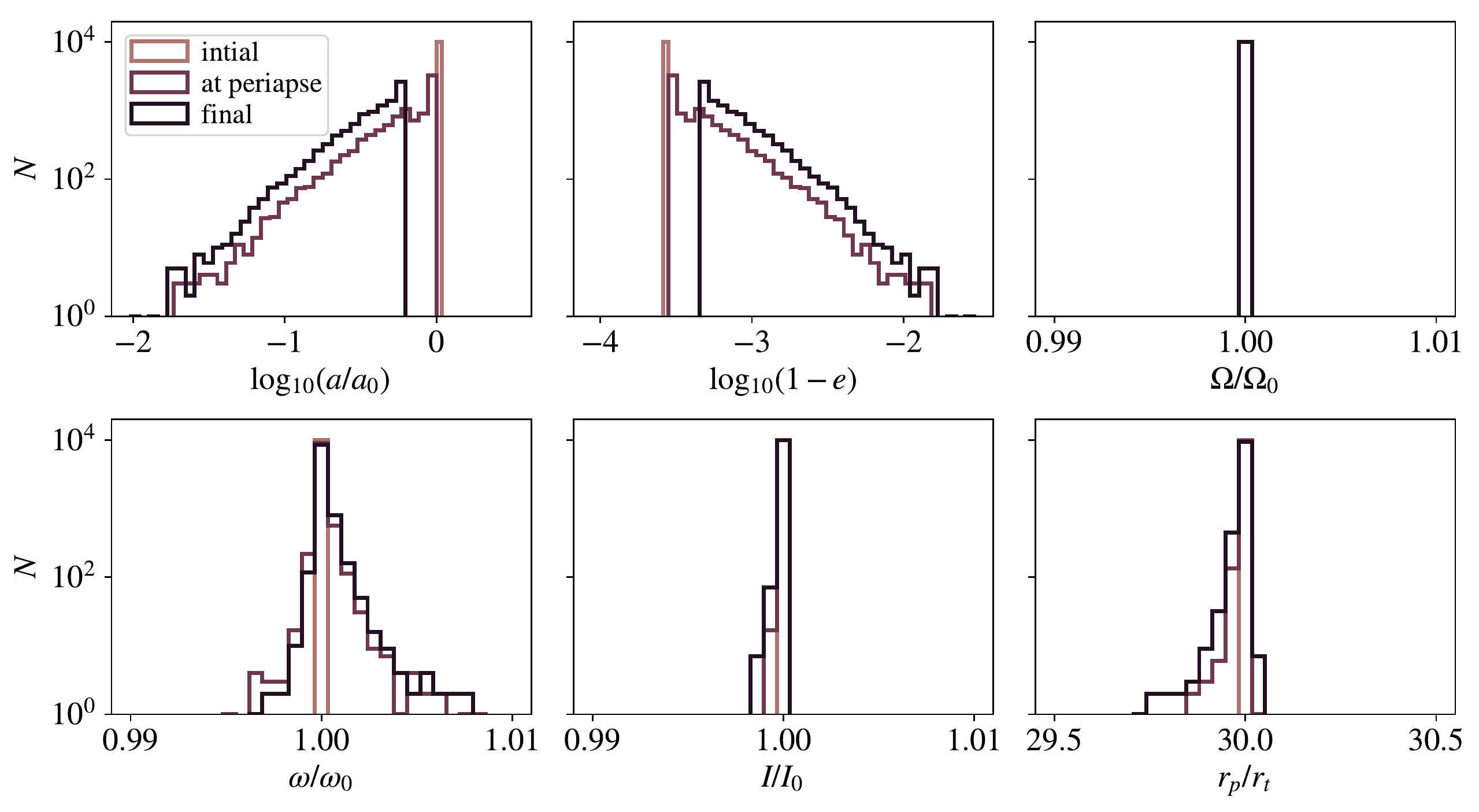}
\caption{Modifications of orbital elements arising after a single orbit's disk crossings. In this example, we adopt an $\alpha=\lambda=Q=1$  disk (ie a gravitationally unstable disk accreting at the Eddington mass accretion limit) around a $10^7 M_\odot$ black hole.  As initial conditions we adopt a single orbital semi-major axis, and eccentricity (implying periapse distance $r_p=30r_t$), but randomize over orbital orientation relative to the disk The panels here show the resultant distributions of orbital elements at apoapse (our initial condition), after a single, ``inbound" disk crossing (labeled ``at periapse"), and after the second, ``outbound" disk crossing (labeled ``final"). Orbital orientation, as parameterized by $\Omega$, $\omega$, and $I$, is relatively unchanged by disk crossings, but orbital semi-major axis (and as a result eccentricity) decreases dramatically.   }
\label{fig:elementsdist}
\end{center}
\end{figure*}

To illustrate how eccentric orbits are modified by drag forces as they pass through the disk, we select a representative sun-like star ($\Ms=M_\odot$ and $\Rs = R_\odot$) and place it on in orbit in which its periapse distance is 30 times the tidal disruption radius,
\beq
\rt =  \left(\frac{\Mbh}{\Ms} \right)^{1/3} \Rs. 
\eeq
If we write $\rp = \beta^{-1} \rt$, then $\beta = 1/30$ for this orbit.  We select an initial semi-major axis inside the black hole's sphere of influence by a factor of 10, such that $a_0=0.1 \rh \approx 0.52$~pc. Together, properties imply an initial eccentricity of $e\approx 0.9997$.  Given these fixed orbital dimensions, we randomize the orientation relative to the disk through generating $10^4$ samples of  the orbital elements $\Omega$, $\omega$, and $I$.  We note that because the system is axisymmetric, we expect outcomes to be independent of $\Omega$. From the initial random distribution, we remove any orbits that are inclined such that they pass within a scale height of the disk plane.  Because $h$ is an increasing function of $r$ as seen in Figure \ref{fig:disk}, we evaluate this condition at apoapse, removing orbits that satisfy $|\tan I | < h(r_{\rm apo})$, where $r_{\rm apo}=a(1+e)$.   

Figure \ref{fig:elementsdist} examines the distributions of orbital elements that result from a single orbital cycle given initially isotropic orientations. We imagine the orbit starting at apoapse, passing through the disk once on the way to its periapse passage, and again after periapse (as shown in Figure \ref{fig:diagram}). Figure  \ref{fig:elementsdist} labels the corresponding distributions of orbital elements initial, at periapse (after one disk crossing), and final (after the second disk crossing). 

In Figure \ref{fig:elementsdist}, we observe that the orientation this eccentric orbit is essentially unmodified by the disk crossings. The angles $\Omega$, $\omega$, and $I$, all remain within 1\% of their initial quantities.  By contrast, the semi-major axis and eccentricity both decrease significantly. The peripase distance $\rp$, is mildly affected, with most orbits showing slightly smaller $\rp$, and a few acquiring larger $\rp$. 

We can qualitatively understand the relative magnitude of these changes in terms of the orbital and disk-crossing geometry. Disk crossings occur very close to periapse for an eccentric orbit. As a result, any drag force applied can significantly modify the orbital energy (this manifests as a decrease in orbital semi-major axis and eccentricity). However, the torques applied to the orbit are relatively small because of the short lever arm of the small periapse distance, leading to comparatively minor changes in the orbit's vectorial angular momentum, equation \eqref{eq:Jvec}. Therefore the angular momentum's  magnitude, which determines $\rp$, and direction, which determines $\Omega$, $\omega$, and $I$, both experience little modification.  

\citet{1995MNRAS.275..628R} showed that as orbits become less eccentric ($e\lesssim 0.9$), the angular torques become comparable to the orbital energy dissipation. This results in a decrease in orbital inclination relative to the disk as an orbit circularizes over subsequent passages. Thus, while orientation is initially unaffected while the orbit is still eccentric, it eventually decreases and stars are entrained into the disk plane if they fully circularize.

\begin{figure*}[tbp]
\begin{center}
\includegraphics[width=0.46\textwidth]{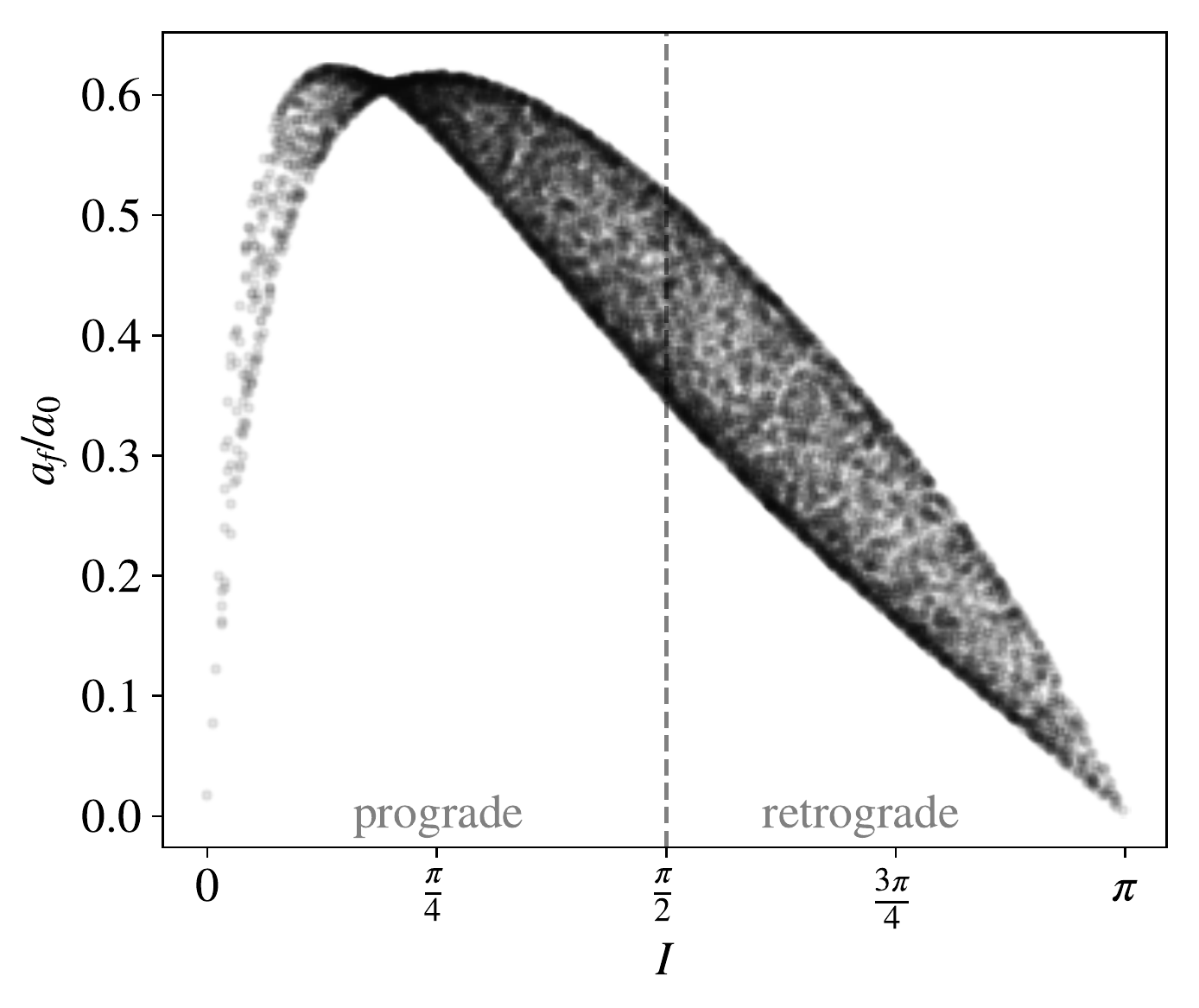}
\includegraphics[width=0.48\textwidth]{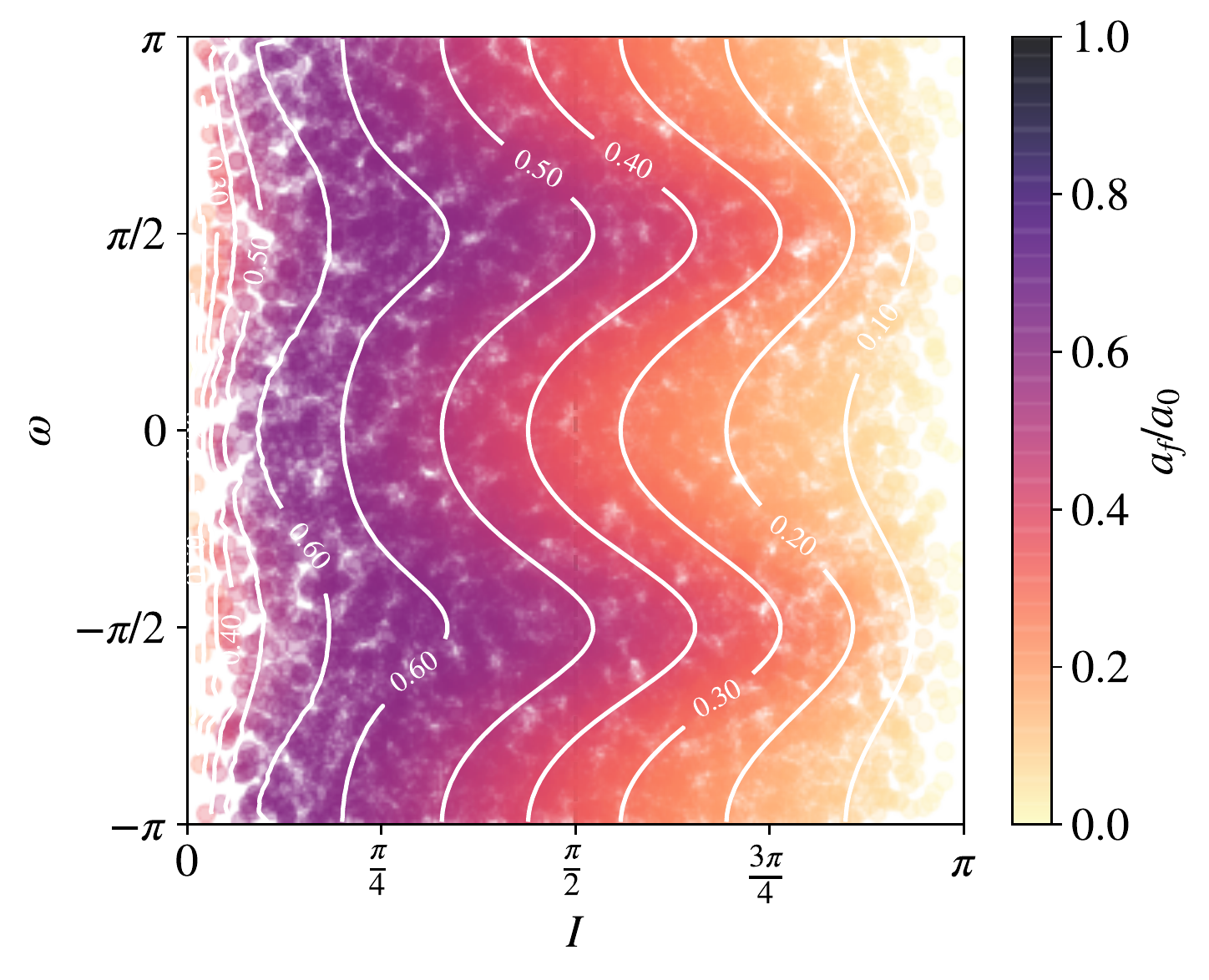}
\caption{The dependence of changes in orbital semi-major axis relative orbit-disk alignment. Here we analyze the same distributions as in Figure \ref{fig:elementsdist}. The left panel highlights the dependence on orbital inclination, while the right panel examines the distribution in inclination and argument of periapse phase space.   Inclinations near $I\rightarrow0$ and $I\rightarrow\pi$ imply orbits that are close to the disk plane, these mildly inclined orbits experience extended disk crossings and the most dramatic changes in orbital semi-major axis.  Retrograde orbits experience more dramatic changes than prograde because of the larger star-disk relative velocity that this configuration implies. By comparison to the dramatic dependence of inclination, argument of periapse, $\omega$, has a minor effect on the amplitude of orbital tightening,  which is maximized at $\omega\rightarrow \pm \pi$ and 0.    }
\label{fig:elementsIomega}
\end{center}
\end{figure*}

One clear conclusion from Figure \ref{fig:elementsdist} is that a distribution of outcomes are realized depending on the orbital orientation. focusing on the ratio of final-to-initial orbital semi-major axis, $a_f/a_0$, Figure \ref{fig:elementsIomega} illustrates some of the dependence of outcome on orientation. We see that orbits that near the disk plane, $I\rightarrow 0$ and $I\rightarrow \pi$, experience most dramatic effect from disk interaction. These trajectories have low $\hat Z$ velocities, and therefore pass most slowly through the disk material (equation \ref{eq:tcross}), and, as a result, traverse the most significant column of disk mass each orbit. Among out of plane orbits, those that are retrograde generally experience greater reduction of semi-major axis. This is attributable to the larger relative velocity between the disk gas and the star that result from retrograde motion, and the resultingly larger drag force in the geometric limit.

Figure \ref{fig:elementsdist} also highlights the dependence in $a_f/a_0$ on argument of periapse, $\omega$. We see that the reduction of semi-major axis is maximized when $\omega \rightarrow \pm \pi/2$. In these orientations, the two disk crossings are roughly equidistant from the black hole, maximizing the amount of dense inner disk material that the star intercepts. As $\omega\rightarrow 0$ or $\pm \pi$, the orbit and disk geometry is such that there is one disk crossing near periapse and one near apoapse. As expected, we do not observe dependence on $\Omega$, which amounts to azimuthal rotation relative to the axisymmetric disk.

\subsection{Comparison to Two-Body Relaxation}\label{sec:twobody}

The preceding analysis has shown that highly eccentric orbits can be modified, in some cases dramatically, by their interaction with the accretion disk. Here we compare the magnitude of these disk-related changes to those that arise from two-body scattering of stellar orbits, which occurs continuously in galactic nuclei as stars trace orbits influenced by both the black hole and the surrounding stellar cluster. 

\begin{figure}[tbp]
\begin{center}
\includegraphics[width=0.99\columnwidth]{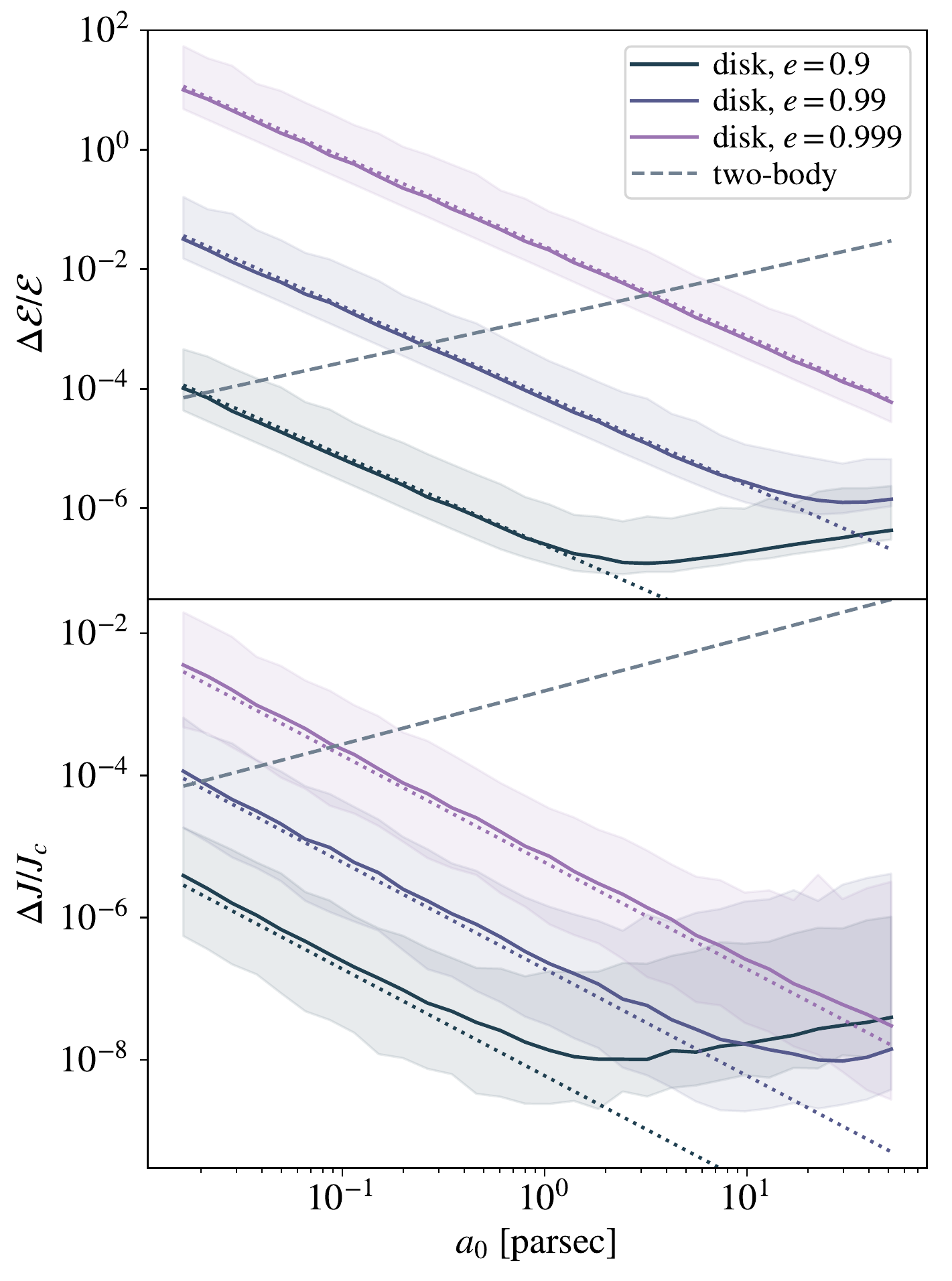}
\caption{Comparison of disk interaction and two-body relaxation for a range of orbital semi-major axes and eccentricities. We adopt the same disk model and $10^7 M_\odot$ black hole as in Figure \ref{fig:elementsdist}. Dashed lines show the RMS fractional changes in energy and angular momentum from two-body relaxation. Shaded regions comprise the 5th to 95th percentile distributions of outcomes depending on relative disk-orbit inclination. Solid lines represent the median result. Finally, the dotted lines compare the to analytic estimates of Section 3.3 for the geometric cross section limit.  At mild eccentricities, $e\lesssim0.9$, two-body relaxation dominates stellar orbital evolution at most orbital semi-major axes. At higher eccentricities, stellar orbits within the black hole's sphere of gravitational influence $(a_0 \lesssim {\rm pc})$, experience much greater influence from disk interaction than from two-body relaxation.  }
\label{fig:twobody}
\end{center}
\end{figure}

Figure \ref{fig:twobody} compares the RMS changes in orbital energy and angular momentum from two body relaxation (per orbit, equations \eqref{eq:DeltaErms} and \eqref{eq:DeltaJrms}, respectively) to the impact of the disk. For disk quantities we show the median value and the 5\% - 95\% region of results, marginalizing over orbital orientation. We show orbits of three representative eccentricities and a range of semi-major axes relative to the black hole sphere of influence radius, $\rh$. 

At fixed eccentricity, the most compact orbits (smallest $a_0$) have the largest $\Delta \E / \E$ and $|\Delta J| / J_c$. These orbits also have the smallest periapse distances and pass through the highest-density inner regions of the disk. As previously discussed,  the magnitude of relative changes to orbital energy from the disk is larger than those to orbital angular momentum. The change in slope seen in the various disk lines occurs due to the transition from geometrically dominated drag cross section (small $a_0$) to gravitationally dominated drag cross section (large $a_0$, where the orbital velocity is lower). 

Figure \ref{fig:twobody} implies that for most of the phase space of stellar orbits, two body relaxation is the dominant dynamical mechanism driving the evolution of orbital energy and angular momentum relaxation. For orbits within the black hole sphere of influence,  $a_0 \lesssim \rh \sim 5~$pc,  the disk becomes a significant driver of orbital energy change when orbits reach eccentricities $e\gtrsim0.99$. By comparison, the disk rarely dominates orbital angular momentum evolution.

\subsection{Order of Magnitude Interpretation and Scalings}\label{sec:oom}

To give a sense of the approximate form of the results reported above and their scalings to other orbital, black hole, and accretion disk properties, we derive order of magnitude estimates here for the disk changes to orbital energy and angular momentum.  For related derivations, see \citet{1991MNRAS.250..505S} and Section 2.2 of \citet{ 2005ApJ...619...30M}. 

The disk mass intersected by a crossing star is key in estimating the magnitude of changes to orbital energy and angular momentum. For the highly eccentric orbits we consider here, the geometric cross section is typically appropriate, and, therefore the disk mass intersected in a passage is
\beq\label{eq:Deltamsimple}
\Delta m \approx \pi \Rs^2 \Sigma,
\eeq
where we have adopted the simplification of the star's orbit passing perpendicular to the disk plane.  Thus, a key quantity to consider is the stellar mean surface density $\Ms/\Rs^2$, as compared to $\Sigma$. Main sequence stars of different masses or stars at differing evolutionary states will have different $\Ms/\Rs^2$, as we will consider below. Expanding equation \eqref{eq:Deltamsimple} in terms of the disk surface density, we find
\beq\label{eq:Deltam}
{\Delta m \over \Ms} \approx {h\over Q}\left(\frac{\Rs}{\rp} \right)^2 \left(\frac{\Mbh}{\Ms} \right),
\eeq
where $\rp$ is the periapse radius of at which the star crosses the disk. Substituting representative parameters for a thin disk and periapse distance of $1000\Rs$,
\beq
{\Delta m \over \Ms}  \approx 10^{-3} \left( \frac{h/Q}{10^{-3}} \right)  \left( \frac{\Rs/ \rp}{10^{-3}} \right)^2   \left( \frac{\Mbh/\Ms}{10^{6}} \right). 
\eeq
This expression implies that when the periapse distance is relatively small, the star can intersect a significant fraction of its own mass in a single disk crossing.  

We can re-write the periapse distance in terms of the tidal disruption radius for stars of a given mass and radius,
\beq
\rp = \beta^{-1} \rt =  \beta^{-1} \left(\frac{\Mbh}{\Ms} \right)^{1/3} \Rs,
\eeq
where the dimensionless impact parameter is $\beta = \rt / \rp$. Substituting this in to equation \eqref{eq:Deltam} to parameterize the disk mass intersected by stars with periapse distance a fixed multiple of their tidal disruption radius, we have
\beq\label{eq:Deltambeta}
{\Delta m \over \Ms} \approx {h\over Q} \beta^2 \left(\frac{\Mbh}{\Ms} \right)^{1/3}.
\eeq
Therefore, in this form, we note that the intersected mass scales with the square of dimensionless impact parameter, $\beta$, (becoming larger as the periapse distance becomes smaller) and only weakly with the ratio of black hole to stellar mass.

From equation \eqref{eq:DeltaV}, we can estimate the change in orbital energy and angular momentum that results from a disk crossing. To do so we make several approximations. We imagine that the star crosses the disk only once, at periapse. We further approximate the relative velocity between the star and the disk to be the periapse velocity (equivalent to disk gas that is at rest rather than orbiting), and assume that the crossing is perpendicular to the disk plane. 
Given these approximations, the magnitude of the change in specific energy is,
\beq\label{eq:DeltaE}
\Delta \E \approx  2H  {F_{\rm drag} \over \Ms} =  {\Delta m \over \Ms}  v_{\rm p}^2 , 
\eeq
where $2H$ is distance traversed as the star crosses the disk and where we have used a periapse velocity, $v_{\rm p} \approx \sqrt{2 G \Mbh / \rp}$, for a highly eccentric orbit. Similarly, the change in specific angular momentum is,
\beq\label{eq:DeltaJ}
\Delta J \approx -  {F_{\rm drag} \over \Ms}   {2H \over v_{\rm p}} \rp = - {\Delta m \over \Ms}  \rp v_{\rm p}, 
\eeq
where $2H/v_{\rm p} \approx \Delta t_{\rm cross}$.
The fractional change in specific energy  is 
\beq\label{dEoE}
{\Delta \E \over \E} \approx 4 {\Delta m \over \Ms} {a \over \rp} = 4 {\Delta m \over \Ms} (1-e)^{-1}.
\eeq
Here the positive sign indicates that orbits become more bound following interaction with the disk (higher $\E$). The fraction change due to loss of specific angular momentum is 
\beq\label{dJoJ}
{\Delta J \over J} \approx  - {\Delta m \over \Ms}.
\eeq

In Figure \ref{fig:twobody} we apply equations \eqref{eq:DeltaE} and \eqref{eq:DeltaJ}, respectively (plotted with dotted lines). We see that these simple expressions provide an excellent description of the median value of the numerical results, provided that the  passage is in the geometrically dominated limit (as assumed in equation \ref{eq:Deltam}).  We also note here that comparison to Figure \ref{fig:twobody} gives a sense of the spread about the median due to differing orbital orientations.

The preceding analysis indicates that for very eccentric orbits, the fractional change in orbital energy is always larger than the fractional change in orbital angular momentum, by a factor $a/\rp = (1-e)^{-1}$. 
This implies that orbits will become more circular under the influence of crossing the disk at periapse, as we have previously observed with our numerical approach.  

We can observe from equations \eqref{eq:Deltamsimple} and \eqref{eq:Deltambeta} that the magnitude of the disk effect depends linearly on the disk surface density (or, equivalently $h/Q$) and on the inverse-square of periapse distance relative to the given star's tidal radius. Because disk surface density depends on our model parameters as $\Sigma \propto \alpha^{-1/3} \lambda^{1/3} Q^{-2/3}$, the mass intercepted is proportional to $\dot M$, or equivalently the fraction of Eddington mass accretion rate $\lambda$. 

More massive black holes will have larger ratio $\Mbh/\Ms$. Equation \eqref{eq:Deltambeta} shows that this yields mildly larger intercepted masses from disk-crossings at fixed $\beta$, but the relaxation time \eqref{eq:trel} is also longer in stellar clusters surrounding more massive black holes. This indicates that disk crossings are more likely to be of significance in systems with high black hole masses than in systems with low black hole mass relative to other stellar dynamical processes.

\subsection{Compact Objects}\label{sec:co}
When the stellar-mass object is a compact object rather than a star, its geometric cross section is reduced accordingly with its compact radius. For illustration, let us take the case of a stellar mass black hole, for which the gravitational focus cross section will dominate for all orbital velocities (because the relative star--disk velocity is always less than the speed of light). In this limit, we find that the magnitude of changes to orbital properties is very small, even in relatively extreme configurations of close-in or very eccentric orbits.  The simplest way to illustrate this is following the order of magnitude scalings of the previous subsection. 

In the gravitational-focus limit, the intercepted disk mass is
\beq
\Delta m \approx \pi R_{\rm a}^2 \Sigma,
\eeq
where $R_{\rm a} = 2 G M_\ast / |\vec V_{\rm rel}|^2$. If we adopt $|\vec V_{\rm rel}| \approx v_{\rm p}$, then 
\beq\label{dMoM_gf}
\frac{\Delta m}{M_\ast} \approx \frac{M_\ast}{M_{\rm bh}} \frac{h}{Q},
\eeq
where we note that there is no dependence on $r$.  The implied fractional change in orbital energy can be derived from equation \eqref{dEoE} and the fractional change in angular momentum from equation \eqref{dJoJ}. Equation \eqref{dMoM_gf} implies that $\Delta m / M_\ast \ll 1$ for compact objects because $M_\ast / \Mbh \ll 1$ and $h/Q < 1$. As a result, regardless of their orbital configuration, compact objects experience only very minor fractional changes in the orbital configurations due to drag forces during passages through the accretion disk.

\section{Potential Consequences for Star--Disk Coexistence}\label{sec:stardisk}

In this section, we discuss some potential implications of our results for the coexistence of dense stellar clusters and accretion disks in galactic nuclei.

\subsection{Stripping and Transformation of Eccentric Stars}

The fact that $\Delta m / \Ms$ can be large when orbits become quite eccentric implies that the star is being forced through a column of mass similar to its own. Additionally, for highly eccentric orbits, the periapse velocity of the orbit (and thus the star--disk relative velocity) tends to be much larger than the star's escape velocity. What does this imply for the star itself? 

During the disk passage, the outer layers of the star endure hydrodynamic drag and shock heating. 
\citet{1993ApJ...407..588M} explored hydrodynamic ablation of self-gravitating spheres, while \citet{1996ApJ...470..237A} and \citet{2016ApJ...823..155K} have performed hydrodynamic simulations of giant stars intersecting columns of material. These numerical results suggest that the momentum imparted in the passage is crucial in the eventual mass removal, as is shock-heating of the envelope material. However, whereas the dissipated energy may be radiated away, the total momentum is conserved; for simplicity, we focus on the momentum transfer in what follows.  

One simple model for momentum transfer can be derived if we assume that a fraction $\eta<1$ of the dissipated orbital momentum goes into stripping mass from the  outer envelope of the star, while the remainder goes into slowing its bulk motion. In this case, the momentum imparted to removing mass from the star at the stellar escape velocity is
\beq
\Delta M_\ast v_{\rm esc}  \sim \eta \Delta m |\vec V_{\rm rel}| \sim \eta \Delta m v_p,
\eeq
where we have approximated the star-disk relative velocity as the periapse velocity. Thus, 
\begin{align}
{\Delta M_\ast \over M_\ast } &\sim \eta  \frac{v_{\rm p}}{v_{\rm esc}} {\Delta m \over M_\ast} ,
\end{align}
where $\Delta m/M_\ast$ is given in Section \ref{sec:oom}, and where $v_{\rm esc} =\sqrt{2 G M_\ast / R_\ast}$, while $v_{\rm p} = \sqrt{2 G M_{\rm bh} / r_{\rm p} }$.  

How does the rate of orbital circularization compare to the rate of mass stripping?  To entirely circularize an orbit such that $e\rightarrow 0$, the star must pass through a column of mass similar to its own mass $\Delta m / M_\ast \sim 1$, equation \eqref{eq:DeltaE}.  For the star's envelope to survive this process ($\Delta M_\ast /M_\ast<1$)  without being removed requires $\eta v_{\rm p} / v_{\rm esc} < 1$. Though the value of $\eta$ is, a priori, unknown, these scalings indicate a dependence on periapse distance through the ratio of stellar escape velocity to peripase velocity. We can, for example, re-express this condition in terms of the stellar tidal radius, 
\beq\label{eq:striptidal}
{\rp \over \rt} > \eta^2  \left( \frac{M_{\rm bh}}{M_\ast}\right)^{2/3}.
\eeq 
For example, if $\eta = 0.1$, then stars passing with $\rp \lesssim 460 \rt$ would have their envelopes ablated rather than fully circularized. 

 The fate of a star under circumstances where envelope ablation occurs likely depends on the star's internal structure and its response to mass loss. Stars with highly differentiated core versus envelope densities may have their envelopes removed while a denser core remains. These denser cores would pass through smaller columns of disk mass and, depending on the precise conditions, might not circularize or ablate significantly. By contrast, a more homogeneous star, like a lower-mass main sequence star, might be fully disrupted if ablative processes are important. The object's response to mass loss is significant here too in the context of repeated disk passages. If, upon losing mass, the star contracts such that its surface density, $\Ms/\Rs^2$, decreases, it will intercept a smaller column of disk mass. However if the star, or its outer envelope, has lower surface density after mass loss, a runaway of stripping (or orbital circularization) would be expected over the course of subsequent orbits.  

\citet{1996ApJ...470..237A} and \citet{2016ApJ...823..155K} have specifically focused on the case of red giants and the implications of these mass-stripping episodes. Because of their extended radii, giants interact with mass similar to their own in orbits with less extreme periapse radii (or lower eccentricities).  Equivalently, we can note that their lower surface density is comparable to the disk surface density at larger radii. Finally, the composite core-envelope structure of these stars implies that even if the hydrogen envelope is removed, the white dwarf core will remain. \citet{1996ApJ...470..237A} estimate a critical radius for the entire giant envelope to be removed over the giant branch lifetime and find a semi-major axis of approximately $0.05$~pc under typical assumptions for quasar disks, and argue that for giants the stripping process is much more efficient that trapping within the disk ($\eta \sim 0.05$ if equation \eqref{eq:striptidal} is applied). Following up on \citet{1996ApJ...470..237A}'s work, \citet{2016ApJ...823..155K} consider the case of our own galactic center and focus on more compact giant-branch stars. They find that collisions with overdense clumps contribute significantly to  stripping these giants if the number of clumps in (or equivalently total mass of) a fragmenting gas disk is high enough -- approximately several hundred times the mass of the current young stellar disk observed in the galactic center, or several $10^4M_\odot$. The paucity of observed giants in the galactic center is  indicative that such a process may have been at play.

\subsection{The Disk Loss Cone: Depletion of Eccentric Orbits}\label{sec:disklc}

\begin{figure}[tbp]
\begin{center}
\includegraphics[width=0.99\columnwidth]{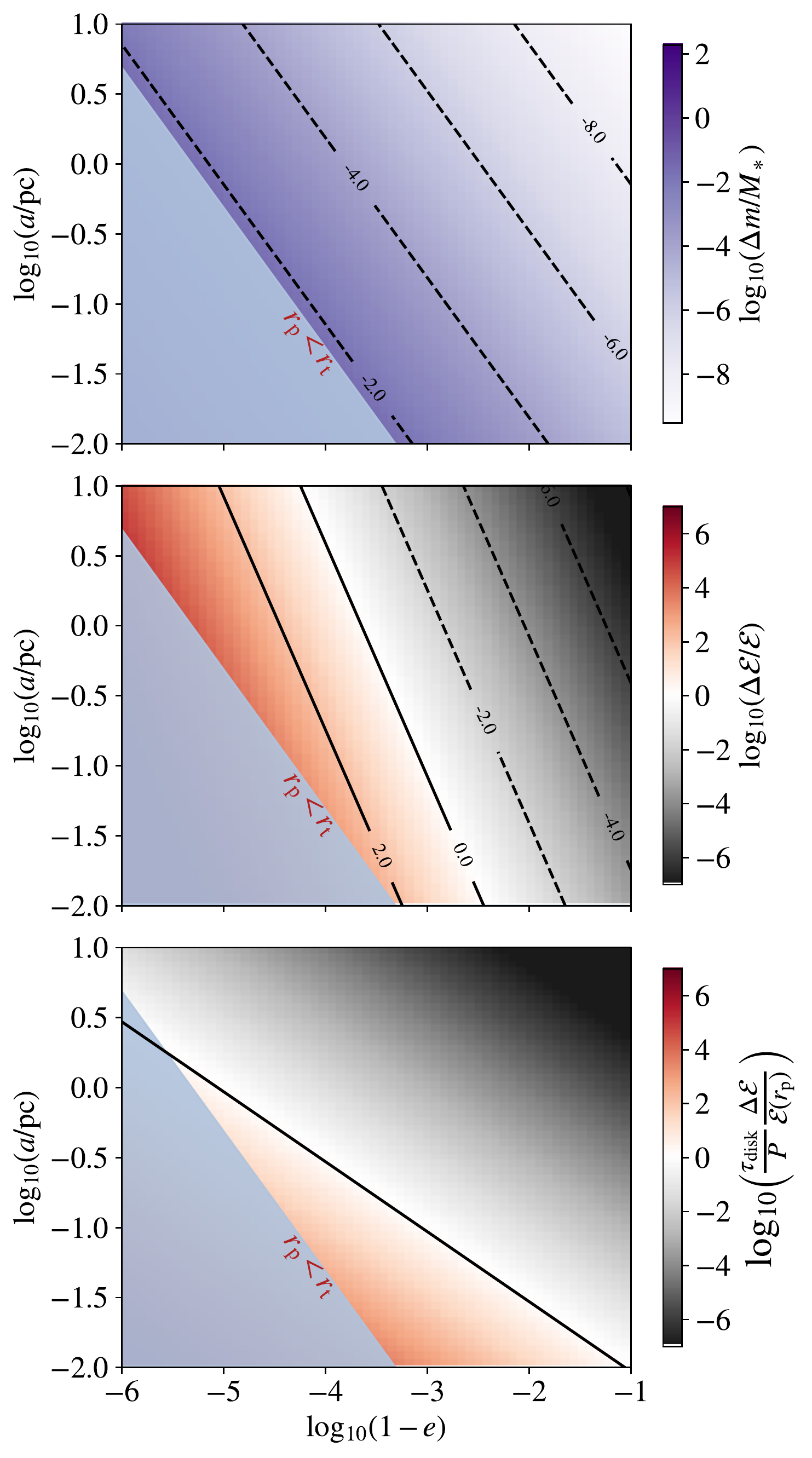}
\caption{The ``loss cone" defined by star-disk interactions. The upper panel shows the mean fractional disk mass intercepted,  $\Delta m /M_\ast$, equation \eqref{eq:Deltam}, for sun-like stars on orbits of varying semi-major axis and eccentricity. In each panel, the tidal-disruption loss cone region of phase space is shaded grey and labeled $\rp < \rt$.  The center panel shows the per-orbit mean fractional change in orbital energy. The lower panel compares the accumulated change in orbital energy over a disk lifetime $\tau_{\rm disk}=1$~Myr, compared to the energy of a circular orbit at that semi-major axis. In regions where $\log_{10}\left( \frac{\tau_{\rm disk}}{P} \frac{\Delta \E}{\E_{\rm c}}\right)>0$, the accumulated dissipation of orbital energy and angular momentum is sufficient to circularize the stellar orbit over the disk lifetime. These orbits are, therefore, lost from the nuclear-cluster phase space over the course of the active accretion phase onto the central black hole.  }
\label{fig:lc}
\end{center}
\end{figure}

In the previous sections, we have shown that the primary effects on stars in highly eccentric orbits are the circularization of their orbits and hydrodynamic ablation due to accumulated passages through the disk near periapse.  
Either of these actions removes stars from the phase space of highly eccentric orbits within the nuclear stellar cluster. Here we examine this depletion of stellar orbits in more detail. 

Figure \ref{fig:lc} defines a ``loss cone" of low angular momentum phase space in which orbits are effectively depleted by disk interaction.  The upper panel of Figure \ref{fig:lc} shows the approximate disk mass intercepted by a sun-like star on orbits of varying semi-major axis and eccentricity, equation \eqref{eq:Deltam}. Here we have again assumed our fiducial disk model of Section \ref{sec:results}, a $10^7M_\odot$ black hole and an $\alpha=Q=\lambda=1$ accretion disk.  In the center panel we compare the per-orbit dissipation of specific orbital energy to the orbital energy, $\Delta \cal E / \cal E$, estimated from equation \eqref{dEoE}.  Finally, the lower panel defines a ``loss cone" of orbits that are circularized over the disk lifetime. The critical condition is $\Delta \E_{\rm total} / \E(r_{\rm p})=1$, where $\Delta \E_{\rm total} = N_{\rm orb}\Delta\E = \Delta \E \tau_{\rm disk}/ P $ is the net accumulated orbital energy dissipated over the disk lifetime, assumed to be $\tau_{\rm disk}=1$~Myr in Figure \ref{fig:lc}. From equation \eqref{dEoE},  $\Delta \E_{\rm total} / \E(r_{\rm p})$ is related to the disk mass intercepted by   $\Delta \E_{\rm total} / \E(r_{\rm p}) \approx  4 {\Delta m / \Ms}$. 

It is useful to compare the disk loss cone to the tidal-disruption loss cone defined by $\rp < \rt$. What we observe from Figure \ref{fig:lc} is that, for black holes in extremely high accretion states, the disk loss cone is larger (extends to lower eccentricity) than the tidal disruption loss cone, especially for orbits with semi-major axis less than a parsec.  The disk-interaction loss cone is especially extended for tight orbits (and has a different slope in phase space than the tidal loss cone) because tight orbits have shorter orbital periods that imply more cumulative passages through the disk over the disk lifetime.

\begin{figure*}[htbp]
\begin{center}
\includegraphics[width=0.95\textwidth]{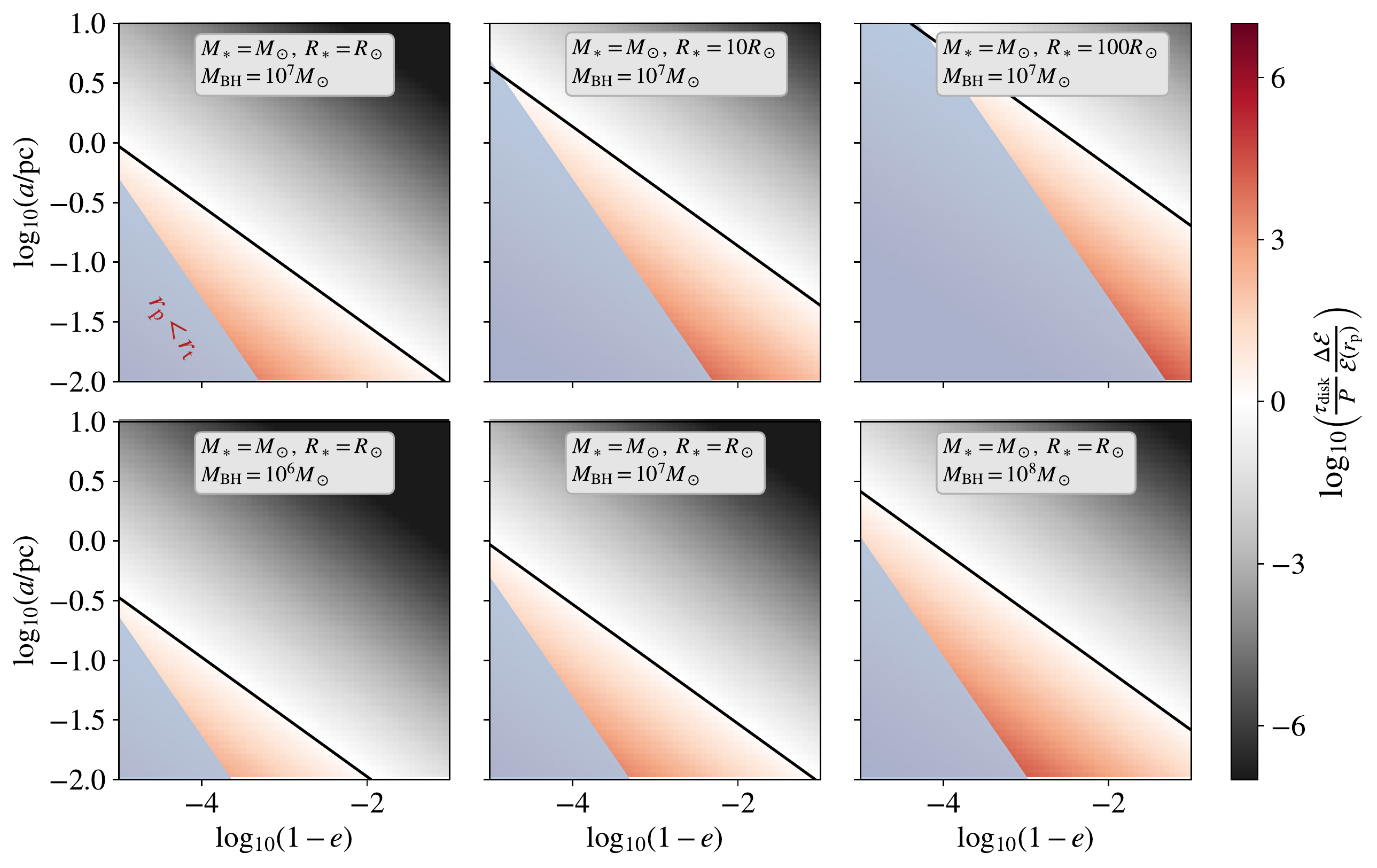}
\caption{Variations in the star-disk ``loss cone"  with increasing stellar radius (top row), and increasing black hole mass (bottom row). All models adopt an $\alpha=\lambda=Q=1$ gravitationally unstable disk accretion at the Eddington mass accretion limit.   }
\label{fig:lcvar}
\end{center}
\end{figure*}

In Figure \ref{fig:lcvar}, we compare the relative phase space occupied by the tidal disruption and disk loss cones under varying stellar and black hole parameters. The upper panels assume a $10^7M_\odot$ accreting black hole, and show stars of solar mass and 1, 10, and 100$R_\odot$. With increasing stellar radius, both the tidal disruption and disk interaction loss cones increase in the portion of phase space they occupy. For giant stars of $100R_\odot$, nearly all orbits with $a\lesssim 0.3$~pc are within the disk loss cone. This suggests that disk interaction can efficiently remove giant stars from the innermost regions of a nuclear cluster, as argued by \citet{1996ApJ...470..237A}.  Because, as shown in equation \eqref{eq:Deltambeta}, $\Delta m/\Ms$ is constant with impact parameter in units of the tidal disruption radius, we find that differing stellar properties do not affect the relative configurations of the star and disk loss cones. 

The lower panels of Figure \ref{fig:lcvar} show varying black hole mass for an assumed sun-like star. As the black hole mass increases, the size of the tidal disruption and disk-interaction loss cones increase in tandem, but do not change their relative configuration significantly. Taken together, this implies that during the active accretion phase, there is an enlarged loss cone on the stellar distribution function due to star-disk interaction.

\subsection{Implications for Stellar Tidal Disruption Events}\label{sec:tde}

The presence of the extended disk loss cone in the orbital phase space leads to the question of whether stars are able to penetrate to sufficiently eccentric orbits to be directly disrupted by the black hole's tides. The answer to this question depends on the relative magnitudes of the dissipation by the disk crossings and scatterings by two-body relaxation. Imagine a star undergoing a random walk in angular momentum space. For this star to undergo a tidal disruption event, it must reach a very eccentric configuration (with $\rp < \rt$) without dissipation acting to circularize its orbit instead. In practice, this implies that the per-orbit scatterings $\Delta J$ are sufficiently large that the star can ``jump" across the disk loss cone: from outside (higher angular momentum) than the disk loss cone to inside the tidal disruption loss cone. An analogous dynamical process is at play for stellar-mass compact objects in galactic nuclei, where gravitational radiation's dissipation interacts with two body scattering to form a Schwarzschild barrier to extreme-mass-ratio inspirals \citep{2011PhRvD..84d4024M}. 

\begin{figure}[tbp]
\begin{center}
\includegraphics[width=0.99\columnwidth]{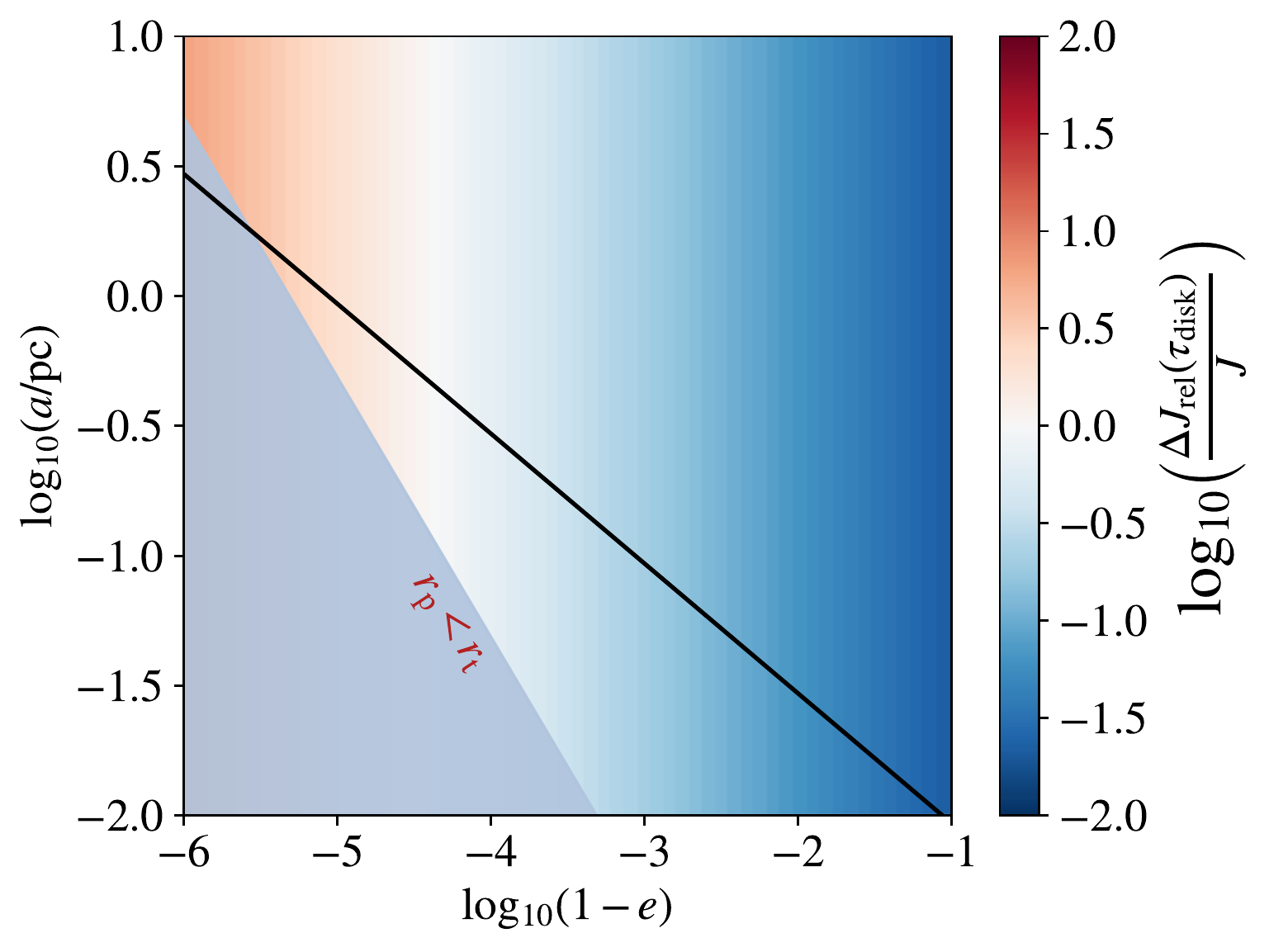}
\caption{The accumulated two-body relaxation driven change in angular momentum over the disk lifetime, compared to the angular momentum at a given $a$ and $e$. Because the fractional change in angular momentum over the disk lifetime is generally less than one, stars are not typically scattered into or out of the disk-interaction loss cone over the disk lifetime. As a result, disk-interaction depletes this phase space of orbits.   }
\label{fig:lcDeltaJ}
\end{center}
\end{figure}

To understand the possible role that two-body relaxation plays in scattering stars into and out of this loss cone phase space, we compare the orbital angular momenta within the disk-interaction loss cone to the random walk in angular momentum accrued over the disk lifetime in Figure \ref{fig:lcDeltaJ}. The accrued random walk over the disk lifetime is proportional to the square root of the number of orbits completed times the per-orbit RMS scatter,  $\Delta J (\tau_{\rm disk}) \approx J_{\rm c} \left( \tau_{\rm disk}/ t_{\rm r} \right)^{1/2} \approx \sqrt{N_{\rm orb}} \Delta J$. For $\nu_\ast \propto r^{-1.5}$, the relaxation time is constant with radius, and $\Delta J (\tau_{\rm disk}) \propto (1-e^2)$. The comparison of this typical scatter to the orbital angular momentum reveals that for all but the most eccentric stars (originating from larger semi-major axes)  stars do not, on average, random walk in or out of the disk-interaction loss cone due to two body relaxation. This finding implies that the phase space of the disk-interaction loss cone in which $\Delta J (\tau_{\rm disk}) < J$  is largely emptied over the course of the disk lifetime, and is not efficiently repopulated by two-body relaxation.

\subsubsection{Flux into the Tidal-Disruption Loss Cone: Default Theory}
To assess the flux of stars into the tidal disruption loss cone with and without the presence of the disk we need to estimate the extent to which two-body relaxation replenishes disk (or tidal disruption) loss-cone orbits. In the simplified case of an isotropic, spherically symmetric stellar distribution function that we have adopted, we can follow the well-developed loss cone theory of the flux of stars into disruptive orbits \citep{1977ApJ...211..244L,1978MNRAS.184...87F,1999MNRAS.309..447M,2005PhR...419...65A,2013degn.book.....M}. 

For an entirely isotropic distribution, the number of stars in the (full) loss cone per unit energy is 
\beq
N_{\rm flc}(\E) = 4 \pi^2 P J_{\rm c}^2 R_{\rm lc} f(\E),
\eeq
where $R\equiv J^2 / J_{\rm c}^2$, and $R_{\rm lc} = J_{\rm lc}^2/J_{\rm c}^2$ is the fraction of the total angular momentum phase space occupied by the loss cone. The angular momentum of the loss cone is 
\beq
J_{\rm lc} = \left[  2 r_{\rm lc}^2 \left( {G \Mbh \over r_{\rm lc}} - \E  \right) \right]^{1/2} \approx (2 G \Mbh r_{\rm lc} )^{1/2}
\eeq
where $r_{\rm lc}$ is the periapse distance of the loss cone. In the case of stars being disrupted by tides,  $r_{\rm lc}= \rt$.   The flux of stars into the loss cone is then $F_{\rm flc} (\E) = N_{\rm flc}(\E) / P$.  

To account for the fact that the loss cone is only partially repopulated by the diffusion process of two-body relaxation, we need to consider the fact that the distribution function depends on angular momentum near the loss cone. A full solution was first derived by \citet{1978ApJ...226.1087C} and is detailed in Chapter 6 of \citet{2013degn.book.....M}, which we follow here. The two-dimensional distribution function is 
\beq\label{fRE}
f(R,\E) = 
\begin{cases}
0, \ \  R \leq R_0 \\
f(\E) \frac{\ln (R/R_0) }{\ln(1/R0) -1 +R_0},  \ \ R > R_0
\end{cases}
\eeq
where $f(\E)$ is given by equation \eqref{fE}. $R_0$ is the dimensionless angular momentum at which the distribution function drops to zero. It is approximately
\beq\label{R0}
R_0 \approx R_{\rm lc} \exp({-\alpha}),
\eeq
where $\alpha = (q^4 + q^2)^{1/4}$, and $q = \Delta J_{\rm rel}^2 / J_{\rm lc}^2$ \citep[equation 6.66]{2013degn.book.....M}. The periapse distance that corresponds to $R_0$ is $r_0 \equiv R_0 a /2 $. 

The number of stars in the loss cone per unit energy can be derived by integration over angular momentum
\beq\label{Nlc}
N_{\rm lc}(\E) = 4 \pi^2 P J_{\rm c}^2   \int_0^{R_{\rm lc}}  f(R,\E) dR,
\eeq
from which the flux of stars into the loss cone is $F_{\rm lc}(\E) = N_{\rm lc}(\E) / P$.  We note that the loss cone flux is often written 
\beq\label{FlcE}
F_{\rm lc}(\E) \approx q \frac{F_{\rm flc}(\E) }{\ln(1/R_0) },
\eeq
which is an approximation of the integral over the angular momentum distribution in equation \eqref{fRE}. Finally, the integrated loss cone flux (the tidal disruption rate of stars) is 
\beq
F_{\rm lc} = \int F_{\rm lc}(\E) d\E  
\eeq
or, explicitly, 
\beq\label{Flc}
F_{\rm lc} = 4 \pi^2  J_{\rm c}^2  \int_{\E_{\rm min}}^{\E_{\rm max}} \int_0^{R_{\rm lc}}  f(R,\E) dR d\E. 
\eeq
Given the power-law distribution function $f(\E)$, the limits of integration $\E_{\rm min}$ and $\E_{\rm max}$ are not particularly well defined. In practice this is mitigated by the fact that $F_{\rm lc}(\E)$ is highly peaked.

\begin{figure}[tbp]
\begin{center}
\includegraphics[width=0.99\columnwidth]{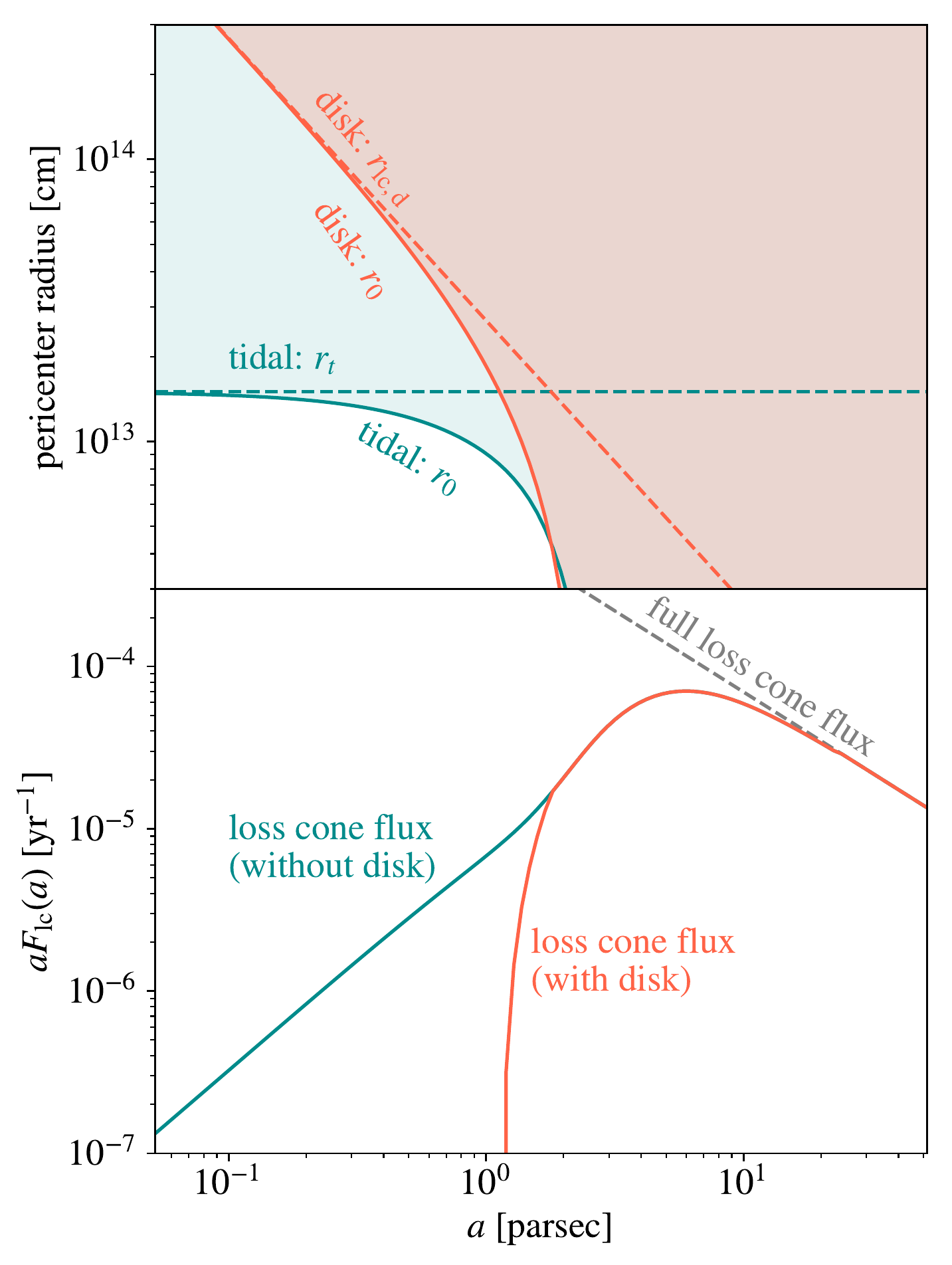}
\caption{Steady-state population of orbits under the influence of two-body scattering and tidal disruption or disk loss cones (top panel) and flux of stars to the tidal disruption loss cone (lower panel). The disk loss cone extends beyond the tidal disruption loss cone, and depopulates orbits with small $a$ from low-angular momentum phase space outside the tidal disruption loss cone. The flux of stars to the loss cone is inhibited at small $a$, but unperturbed at larger $a$ because typical scatterings allow stars to jump over the disk loss cone.  The integrated tidal disruption rate is not significantly impacted.  }
\label{fig:tde}
\end{center}
\end{figure}

Again adopting our fiducial $\Mbh=10^7M_\odot$, $\alpha=Q=\lambda=1$ AGN and cluster models, the upper panel of Figure \ref{fig:tde} compares the loss cone peripase radius for tides $\rt$ to the periapse distance where the distribution function drops to zero, $r_0$, as a function of semi-major axis. Below $r_0$, the phase space is empty. Above $\rt$, it is roughly isotropic. At larger semi-major axes, the per-orbit diffusion in angular momentum is larger, and $r_0\rightarrow 0$ ($q\gg 1$, the ``full loss cone" limit), while at smaller semi-major axes the per-orbit diffusion is weak and $r_0\rightarrow \rt$ ($q\ll1$, the ``empty loss cone" limit).  The lower panel of Figure \ref{fig:tde} shows the flux into the loss cone as a function of semi-major axis. In this particular example, the loss cone flux peaks at $a\sim 5$~pc.

\subsubsection{Modification by Disk Loss Cone}

To consider the role of the disk loss cone, we define equivalent quantities to those of the previous subsection. In particular, we define the disk loss cone periapse distance, $r_{\rm lc,d}$ based on the criterion of circularization over the disk lifetime of section \ref{sec:disklc}, $(\tau_{\rm disk} / P ) \Delta \E  =\E (\rp)$. This peripase radius therefore corresponds to the contours plotted in Figures \ref{fig:lc}, \ref{fig:lcvar}, and \ref{fig:lcDeltaJ}.  The corresponding zero-point peripase distance $r_0$ of the resulting distribution function emerges from the definition $r_0 = R_0 a /2 $, where we now evaluate the $R_0$ on the basis of the disk loss cone in equations \eqref{fRE} through \eqref{Flc}.  

The upper panel of Figure \ref{fig:tde} shows $r_{\rm lc,d}$ and the disk-based $r_0$. We observe that the disk loss cone periapse distance depends on orbital semi-major axis, as expected from Figures  \ref{fig:lc}, \ref{fig:lcvar}, and \ref{fig:lcDeltaJ}. The behavior of the disk $r_0$ relative to $r_{\rm lc,d}$ is qualitatively similar to that described in the tidal disruption case. For small semi-major axes, two-body relaxation does not refill the phase space carved out by disk interaction. For large semi-major axes, $r_0\rightarrow0$ and two-body relaxation leads new stars to diffuse into the disk loss-cone every orbit. 
When $r_0 \ \rt$, then stars diffuse to sufficiently small periapse distances to be disrupted by the black hole. From Figure \ref{fig:tde}, we see that this primarily occurs for $a\gtrsim 1$~pc. By contrast, when $r_0 > \rt$ no tidal disruption events can occur, because stars are lost to disk interaction in their angular momentum diffusion prior to reaching sufficiently low angular momentum to tidally disrupt. 

We can incorporate the disk-modified distribution function, $f_{\rm d} (R,\E)$ into our integration of the stellar tidal disruption flux. In $f_{\rm d}(R,\E)$, the {\it disk} loss cone is used to define $J_{\rm lc}$, $R_{\rm lc}$, and $R_0$.  The number of stars in the tidal disruption loss cone is given by equation \eqref{Nlc}, while the total tidal disruption rate is given by equation \eqref{Flc}, in these expressions, by contrast, the upper integration limit $R_{\rm lc}$ now represents the tidal disruption loss cone. 

Pictorially, comparing to Figure \ref{fig:tde}, we are integrating the stars in the disk-modified distribution function (orange) with periapse distance less than the tidal radius, $\rt$. The lower panel of Figure \ref{fig:tde} shows the disk-modified tidal disruption flux. We see that disk interaction sharply truncates the flux of stars in orbits of $a\lesssim 1$~pc, which diffuse to the tidal disruption loss cone over many orbits; these stars tend to be captured into disk interactions rather than continuing their random-walk in angular momentum. Interestingly, the integrated tidal disruption rate is modified very little, because this cutoff occurs at orbits more tightly bound than the peak of the tidal disruption flux. 

\begin{figure}[tbp]
\begin{center}
\includegraphics[width=0.99\columnwidth]{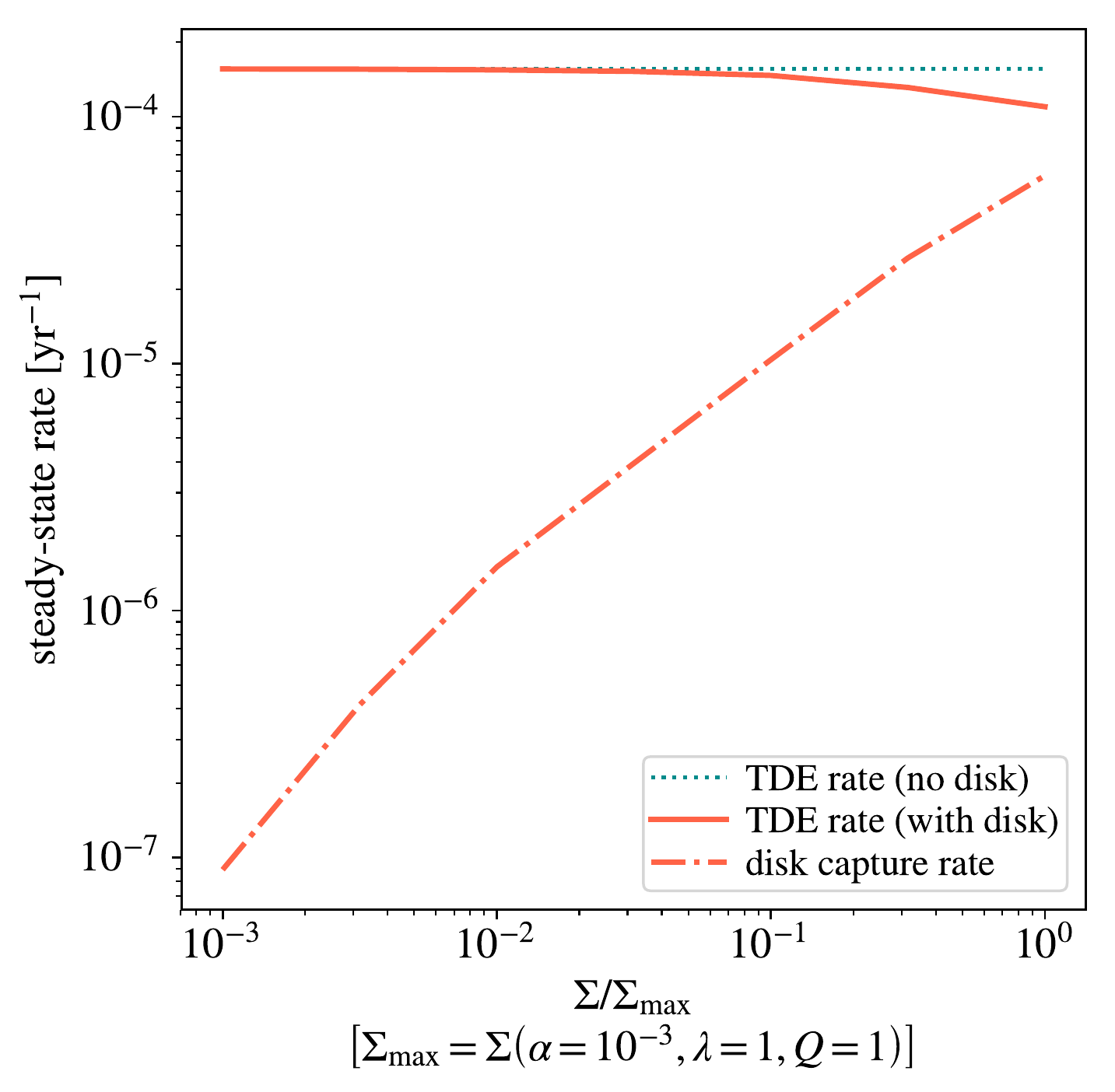}
\caption{Steady-state event rate of tidal disruption events and disk captures in the presence of an AGN disk. Disk properties are parameterized here by the ratio of disk surface density, $\Sigma$, to a maximal model surface density, $\Sigma_{\rm max}$, derived with parameters $\alpha=10^{-3}$, $\lambda =1$, and $Q=1$.  Only as $\Sigma \rightarrow \Sigma_{\rm max}$ is the steady-state tidal disruption event rate decreased by approximately 30\% by losses of stars to the disk loss cone. }
\label{fig:tderate}
\end{center}
\end{figure}

Figure \ref{fig:tderate} shows the integrated tidal disruption event rate with and without a surrounding AGN disk. Here we assume a $10^7 M_\odot$ black hole, surrounded by a stellar cluster as described in Section \ref{sec:nsc}. The presence of the disk loss cone depletes some stellar orbits that would have otherwise diffused to the black hole in the empty loss cone limit. We plot the steady state event rates as a function of the ratio of $\Sigma$ to a maximal $\Sigma_{\rm max}$, which is derived by adopting parameters of $\alpha=10^{-3}$, $\lambda =1$, and $Q=1$.  As $\Sigma/\Sigma_{\rm max} \rightarrow 1$ there is at most a 30\% decrease in tidal disruption event rate for the model parameters we have selected. The diffusion of orbits into the disk loss cone in steady state is also plotted as the ``disk capture rate". This is the rate at which new orbits diffuse into the phase space of the disk loss cone to refill it after its initial depletion due to disk interaction, and is approximately equal to the decrement in the tidal disruption event rate.

We conclude by noting that the steady-state approximation that we have made in adopting the \citet{1978ApJ...226.1087C}  solution to the phase space distribution may not be fully warranted. When stars enter the disk-interaction loss cone, they are brought into more circular orbits, not deleted. Thus, further interaction with the stellar cluster is still possible. A time-dependent solution that follows the combined effects of disk interaction, two-body, and secular relaxation processes is highly worthwhile to pursue to understand the subsequent evolution of captured stars \citep[e.g.][find that many, but not all stars captured by the disk subsequently interact with the black hole by migration or subsequent scatterings]{2016MNRAS.460..240K,2018MNRAS.476.4224P}. 

\section{Summary \& Conclusion}\label{sec:conclusion}

We have examined one aspect of the coexistance of dense stellar clusters and accretion disks in AGN: the role of disk crossings in modifying orbits that become highly eccentric. Highly eccentric orbits cross through the most concentrated inner portions of the accretion disk, and suffer the greatest consequences from these crossings.  Our main findings are:
\begin{enumerate}

\item Disk crossings apply a drag force near the periapse of eccentric orbits (Figure \ref{fig:diagram}). This damping of orbital motion causes highly eccentric orbits to circularize, while initially not dramatically affecting orbital orientation (Figure \ref{fig:elementsdist}). The strongest effects are on stars nearly within the disk plane, which cross through extended columns of disk gas (Figure \ref{fig:elementsIomega}).  As orbits circularize, angular torques become comparable to the decrease in eccentricity, and orbits are incorporated into the disk plane \citep{1995MNRAS.275..628R}. 

\item Drag forces at disk crossings can lead to larger changes in orbital energy and angular momentum than two-body relaxation for orbits that are highly eccentric (Figure \ref{fig:twobody})  and for stars of low mean surface density, $\Ms/\Rs^2$. These stars are likely to decouple from the surrounding cluster's dynamical evolution and evolve primarily through their continued interaction with the disk.  This is particularly true in the case of AGN accreting near the Eddington mass accretion limit. Lower-level AGN disks are not sufficiently dense to dominate stellar orbital histories, even in highly-eccentric configurations; These effects can be seen most clearly through the scalings of Section \ref{sec:oom}. Compact objects experience lower drag forces and are, under typical conditions, always in the stellar relaxation dominated regime as seen in Section \ref{sec:co}. 
 
\item The region of orbital parameter space that leads to stellar capture by the disk may be thought of as a disk ``loss cone" (Figure \ref{fig:lc} and Section \ref{sec:disklc}).  For black holes accreting near the Eddington mass accretion limit, the disk-interaction loss cone can be larger than the stellar tidal disruption loss cone, depleting eccentric orbits that would otherwise diffuse to the supermassive black hole and result in tidal disruption events (Figures \ref{fig:lcvar} and \ref{fig:lcDeltaJ}). 

\item We find that the tidal disruption rate fed by two-body relaxation is relatively unchanged even by the presence of an Eddington mass accretion rate AGN, because stars are still fed to the black hole from more distant orbits with larger-angle two-body scatterings that allow them to ``jump" across the disk loss cone, as discussed in Section \ref{sec:tde} and shown in Figures \ref{fig:tde} and \ref{fig:tderate}.  This finding does not explain the \citet{2017NatAs...1E..61T} result of an elevated tidal disruption event rate in AGN, nor does it rule out the coexistence of tidal disruption events with preexisting accretion flows \citep[e.g.][]{2019ApJ...881..113C}. 

\item Software to reproduce all of the results of this work, and examples that explore further parameter combinations are publicly released accompanying this paper in the form of a python package \href{https://github.com/morganemacleod/NSC_dynamics}{\tt NSC\_dynamics}.

\end{enumerate}

Stars that are entrained within the disk by entering the disk loss cone may undergo one of a number of different fates. Migration within the disk plane may convey some stars to the black hole \citep[e.g.][]{1993ApJ...409..592A,2007A&A...470...11K,2016MNRAS.460..240K}, or may lead to enhanced stellar encounter rates \citep{2016ApJ...819L..17B,2017MNRAS.464..946S,2018ApJ...866...66M,2019ApJ...878...85S}. Stellar evolution continues in the presence of the surrounding disk gas, perhaps leading entrained stars to evolve to compact objects \citep{1993ApJ...409..592A,1994ApJ...423L..19L,1995MNRAS.276..597R}. Stellar dynamics within the emergent stellar disk of entrained stars \citep{1995MNRAS.275..628R,1998MNRAS.293L...1V,2004MNRAS.354.1177S,2018MNRAS.476.4224P} leads to particularly interesting effects, including enhanced star-star encounters, or potentially compact object mergers and their associated gravitational-wave counterparts \citep{2017MNRAS.464..946S}, and enhanced torquing of stars of into highly-eccentric orbits which can lead to tidal disruption events \citep[e.g.][]{2009ApJ...697L..44M,2011ApJ...738...99M,2018ApJ...853..141M,2019ApJ...880...42W} possibly explaining the elevated tidal disruption event rate suggested by the \citet{2017NatAs...1E..61T} observation.

Numerous aspects of the question of star and disk coexistence in galactic nuclei merit continued study. In the particular case of highly-eccentric orbits, we highlight that gas-dynamical models could shed more light on the dissipation of orbital energy and momentum by drag and the stripping of stellar envelopes due to the disk ram pressure \citep{1996ApJ...470..237A,2016ApJ...823..155K}\footnote{However, we acknowledge that this is a numerically challenging problem. The presence of high Mach-number flows and the need to have high spatial resolution in low-mass regions near the stellar atmosphere present challenges to both Eulerian and Lagrangian hydrodynamics techniques.}. Because disk-crossings typically affect stellar orbits over numerous orbital periods, a time-dependent model that includes star-disk interactions and stellar dynamics may also be useful in building a more complete understanding of the ways in which stellar dynamics intertwines with the complex astrophysical environment surrounding an actively accreting black hole. Such work has been undertaken under the assumptions of disk-interactions only \citep{1995MNRAS.275..628R} and more recently with direct N-body calculations with a (necessarily) limited total number of stars \citep{2016MNRAS.460..240K}. As stellar orbits settle within the disk, secular torques become increasingly important \citep{1996NewA....1..149R,2009ApJ...697L..44M,2018ApJ...853..141M} and must be considered to correctly evaluate the star, gas disk, and black hole system's subsequent evolution.

\acknowledgments{
M.M. gratefully acknowledges helpful discussions on this and related topics with A. Antoni, A. Loeb, A.-M. Madigan, M. Rees, and E. Ramirez-Ruiz.  
D.N.C Lin thanks the Institute for Theory and Computation at Harvard University for hospitality and support while a portion of this work was completed. 
M.M. is grateful for support for this work provided by NASA through Einstein Postdoctoral Fellowship grant number PF6-170169 awarded by the Chandra X-ray Center, which is operated by the Smithsonian Astrophysical Observatory for NASA under contract NAS8-03060. 
Support for program \#14574 was provided by NASA through a grant from the Space Telescope Science Institute, which is operated by the Association of Universities for Research in Astronomy, Inc., under NASA contract NAS 5-26555.
This material is based upon work supported by the National Science Foundation under Grant No. 1909203. }

\software{ Astropy \citep{2013A&A...558A..33A}, IPython \citep{PER-GRA:2007}, SciPy \citep{jones_scipy_2001}, matplotlib \citep{Hunter:2007}   }

\clearpage
\bibliographystyle{aasjournal}

\end{document}